\documentclass[twocolumn]{aastex63}

\submitjournal{ApJS}

\shorttitle{Likelihood-free Cosmological Constraints with ANN}
\shortauthors{Wang et al.}

\usepackage{bm}
\usepackage{amssymb}
\usepackage[intlimits]{amsmath}
\usepackage{listings}
\usepackage{python}

\renewcommand{\nolinkurl}{} 
\renewcommand{\url}[1]{\href{#1}{#1}}

\newcommand\tb{\bm{\theta}}
\newcommand\X{\bm{X}}
\newcommand\Hz{$H(z)$ }
\newcommand\z{\bm{z}}
\newcommand\Hb{\bm{H}}
\newcommand\Xb{\bm{X}}

\renewcommand\sb{\bm{\sigma}}
\newcommand\Om{\Omega_m}
\newcommand\Ol{\Omega_\Lambda}
\newcommand\Ok{\Omega_k}
\newcommand\La{$\mathrm{\Lambda}$}

\newcommand\sig{\bm{\sigma}}
\newcommand\obs{_{\mathrm{obs}}}
\newcommand\fid{_{\mathrm{fid}}}

\newcommand\N{\mathcal{N}}

\newcommand\ma{\mathrm{max}}
\renewcommand\L{\mathcal{L}}
\newcommand{\vT}{\bm{\varTheta}}
\newcommand{\yb}{\bm{y}}
\newcommand{\T}{^\mathrm{T}}

\newcommand{\pp}[3][{}]{{\frac{\partial^{#1} {#2}}{\partial {#3}^{#1}}}}

\def\mailto#1{\href{mailto:#1}{#1}}

\newcommand\Papa{Papamakarios}
\begin{document}


\title{Likelihood-free Cosmological Constraints with Artificial Neural Networks: An Application on Hubble Parameters and SNe Ia}
	
	\correspondingauthor{Tong-Jie Zhang}
	\email{tjzhang@bnu.edu.cn}

	\author{Yu-Chen Wang}	
	\affiliation{Department of Physics, Beijing Normal University, Beijing 100875, 
	China}
	
	\author{Yuan-Bo Xie}
	\affiliation{Department of Astronomy, Beijing Normal University, Beijing 100875, 
	China;
\mailto{tjzhang@bnu.edu.cn} 
}
	
	
	\author{Tong-Jie Zhang}
	\affiliation{Department of Astronomy, Beijing Normal University, 
	Beijing 100875, 
	China;
	\mailto{tjzhang@bnu.edu.cn}
    }
    \affiliation{Institute for Astronomical Science, Dezhou University, Dezhou 253023, China}

	\author{Hui-Chao Huang}	
	\affiliation{Department of Physics, Beijing Normal University, Beijing 100875, 
	China}	

	\author{Tingting Zhang}
	\affiliation{College of Command and Control Engineering, PLA Army Engineering University, 
	Nanjing, 
	China; 
	\mailto{101101964@seu.edu.cn}
	}

	\author{Kun Liu}
	\affiliation{College of Command and Control Engineering, PLA Army Engineering University, 
		Nanjing, 
		China;
\mailto{101101964@seu.edu.cn} 
}
	
\begin{abstract}
The errors of cosmological data generated from complex processes, such as the observational Hubble parameter data (OHD) and the Type Ia supernova (SN Ia) data, cannot be accurately modeled by simple analytical probability distributions, e.g. Gaussian distribution. To constrain cosmological parameters from these data, likelihood-free inference is usually used to bypass the direct calculation of the likelihood.
In this paper, we propose a new procedure to perform likelihood-free cosmological inference using two artificial neural networks (ANN), the Masked Autoregressive Flow (MAF) and the denoising autoencoder (DAE). 
Our procedure is the first to use DAE to extract features from data, in order to simplify the structure of MAF needed to estimate the posterior. Tested on simulated Hubble parameter data with a simple Gaussian likelihood, 
the procedure shows the capability of extracting features from data and 
estimating posterior distributions
without the need of tractable likelihood. We demonstrate that it can accurately approximate the real posterior, achieve performance comparable to the traditional MCMC method,  and the MAF gets better training results for small number of simulation when the DAE is added. We also discuss the application of the proposed procedure to 
OHD and Pantheon SN Ia data, 
and use them to constrain cosmological parameters from the non-flat \La CDM model. For SNe Ia, we use fitted light curve parameters to
find constraints on $H_0, \Om, \Ol$ similar to relevant work, using less empirical distributions. In addition, this work is also the first to use Gaussian process in the procedure of OHD simulation.


\end{abstract}

\keywords{Observational cosmology (1146), Neural networks (1933), Computational methods (1965), Astronomy data analysis (1858), Cosmological parameters (339), Astrostatistics strategies (1885)}

\section{Introduction}
\label{sec:intro}

Constraining parameters of cosmological models is one of the basic tasks in cosmology. In recent years, various observational datasets are used to constrain cosmological parameters, including
observational Hubble parameter data \citep[OHD, e.g.][]{Jesus2018}, Type Ia supernovae 
\citep[SNe Ia, e.g.][]{Scolnic2018}, 
cosmic microwave background \citep[e.g.][]{Planck2013}, large-scale structures \citep[e.g.][]{Chuang2017, Pan2019a}, etc. 
Traditionally, analytical likelihood functions are defined to 
describe the probability distribution of the observed data,
then parameters are evaluated by inferring the posterior. 

However, as pointed out in \citet{Weyant2013}, for some datasets the errors of the
data 
can not be perfectly modeled with simple analytical distributions, e.g. multivariate Gaussian distributions.
Thus, the real likelihood of the problems are often complicated
and may not have an analytical form, making it difficult to perform traditional methods to infer parameters. To perform Bayesian inference with an intractable likelihood, several likelihood-free methods have been proposed to bypass the direct calculation of likelihood, e.g. Approximate Bayesian Computation (ABC, see e.g. \citet{Marjoram2003,Bonassi2015a}). ABC has been applied to several astronomical tasks, including SN Ia cosmology \citep{Weyant2013} and galaxy evolution \citep{Cameron2012}. However, 
as another sort of sampling-based method, ABC
gives noisy estimations, and it is sometimes impractical to get high-quality inferences \citep{Papamakarios2016}.

With the rapid development of artificial neural networks (ANN) in the past few decades, several ANN-based likelihood-free inference methods were proposed. These methods evaluate distributions (e.g. the posterior in Bayesian inference) using neural density estimators. Recently, the state-of-the-art density estimator called Masked Autoregressive Flow (MAF) has been proposed by \citet{Papamakarios2017}, who later explored its application in likelihood-free inference \citep{Papamakarios2019}. In several experiments carried out by Papamakarios, MAF was shown to be able to give more accurate estimations of distributions and parameters than similar models.

Despite the excellence of MAF in likelihood-free inference, it may require more simulation and computational resources for higher-dimensional data or irregular posterior distributions. Therefore, reducing the dimensionality of data is essential for efficient training of models. 
Traditionally, this can be done by defining parameters or extracting informative statistics from data heuristically \citep[e.g.][]{Papamakarios2016}. \citet{Charnock2018} proposed using a kind of ANN called "information maximizing neural networks" (IMNNs) to transform data into summaries that approximately satisfy Gaussian distribution, and training the networks to maximize the Fisher information at fiducial values of parameters. The Fisher information involves partial derivatives at the chosen fiducial value with respect to parameters, so it was estimated by simulating data at the fiducial value $\tb^{\mathrm{fid}}$ and near the fiducial value (i.e. $\tb^{\mathrm{fid}}\pm\Delta\tb$) (see \citet{Charnock2018} for details). In an example in \citet{Alsing2019}, IMNNs were used to compress data before carrying out a single-parameter likelihood-free inference using neural density estimators.



We propose extracting data features with another kind of ANN, namely denoising autoencoders 	\citep[DAE,][]{vincent2008, dae},  which can also learn reliable lower-dimensional representations of data corresponding to parameters in the whole possible region. 
As a commonly used dimensionality reduction model, DAE
learns to encode data to encodings such that data can be reconstructed from encodings. Thus, representative features of data can be obtained automatically without artificial choice of statistics. For the parameter inference task in this paper, we also proposed adding another term to the loss function of DAE 
that may
extract even more information about parameters (discussed in Section \ref{sec:ae}).

In this work, we propose a likelihood-free cosmological inference procedure
using DAE and MAF, 
and discuss the application of them in cosmological constraints using 
OHD
and SNe Ia. We also propose using Gaussian Process to better simulate the change of OHD's uncertainty with redshift, and inferring directly from data fitted from SN Ia light curves with less prior assumptions. In Section \ref{sec:model}, we review and discuss in detail the theory and application of the traditional method and ANN methods in cosmological constraints. In Section \ref{sec:simul}, we report the results of the evaluation of DAE and MAF methods on simulated Hubble parameter data, and compare it to the classical MCMC method. To validate the comparison, simple Gaussian uncertainty is assumed so that the likelihood is tractable. In Section \ref{sec:OHD}, we show the results of constraint from real OHD using MAF. In Section \ref{sec:SN}, we discuss the application of DAE and MAF on SN Ia dataset, and perform a preliminary constraint on a cosmological model to describe the proposed procedure. Finally, we conclude and discuss our procedure and results in Section \ref{sec:concl}.


\section{Procedure for Parameter Constraint}
\label{sec:model}
\subsection{Traditional Parameter Inference Method}

To discuss the traditional method for parameter inference, we take the OHD data as an example. The observational dataset is \Hz at given redshifts $\z=(z_1, \cdots, z_N)\T$ and is expressed as $\Hb\obs=\left(H_{\mathrm{obs},1}, \cdots, H_{\mathrm{obs},N}\right)\T$ where $N$ is the dimension of data vector, in other words the number of $H(z)$ points observed.
The free parameters to be inferred from  $\Hb\obs, \z$ are expressed as a $D$-dimensional vector $\tb = \left(\theta_1, \cdots, \theta_d, \cdots, \theta_D\right)\T$, e.g. $\tb=(H_0, \Om, \Ol)\T$ for non-flat \La CDM model. The work of determining $\tb$ is basically estimating the likelihood $P(\Hb\obs|\tb)$ or, with prior knowledge of $\tb$, the posterior distribution of parameters $P(\tb|\Hb\obs)$. 

For OHD, the data is usually modeled with Gaussian distribution, so the parameter inference is traditionally carried out with $\chi^2$ analysis. Assuming that the error $\sig=\left(\sigma_i\right)\T$ of $\Hb\obs$ satisfies a Gaussian distribution with a diagonal covariance matrix\footnote{In principle, there may be covariance between $H(z)$ at different $z$. However, for the 31 existing OHD evaluated with cosmic chronometer method,
there is no estimation of the covariance. Thus, the covariance are usually set to zero, but future work should focus on evaluating the correlations of data.} \citep[as assumed in e.g. ][]{Ma2011}, the likelihood $\L(\tb)$ can be written as
\begin{equation}
\L(\tb) = \label{like}
P(\Hb\obs | \tb)=\left(\prod_{i} \frac{1}{\sqrt{2 \pi \sigma_{i}^{2}}}\right) \exp \left(-\frac{\chi^{2}}{2}\right),
\end{equation}
where the $\chi^2$ statistic is 
\begin{equation}
\label{chi2}
\chi^{2}=\sum_{i} \frac{\left[H(z_{i} ; \tb)-H_{\mathrm{obs}, i}\right]^{2}}{\sigma_{i}^{2}},
\end{equation}
and $H(z_i;\tb)$ is the theoretical Hubble parameter at $z_i$ given a specific set of parameters. For non-flat $\mathrm{\Lambda}$CDM model, 
\begin{equation}
H(z_i;H_0, \Om, \Ol)=H_{0} \sqrt{\Om(1+z_i)^{3}+\Ol+\Ok (1+z_i)^{2}},
\label{eq:fiducial}
\end{equation}
where $\Ok=1-\Om-\Ol$.
With prior knowledge $P(\tb)$, the posterior is given by the Bayes' theorem:
\begin{equation}
\label{bayes}
P(\tb|\Hb\obs) \propto P(\Hb\obs | \tb)P(\tb).
\end{equation}

\explain{RHS of Eq. (\ref{eq:fiducial}) corrected: $z$ changed to $z_i$, to make it consistent with the $z_i$ on the LHS.}

Then, the likelihood or the posterior is evaluated given the aforementioned formulas. Directly computing $\L(\tb)$ on a grid in the parameter space is computationally heavy, and evaluating the marginal distributions of the posterior involves multidimensional integrations. Thus, $\L(\tb)$ is usually estimated by sampling from it with Markov chain Monte Carlo (MCMC) algorithms \citep{Lewis2002, Christensen2001}.\footnote{Apart from the most traditional MCMC, there are other alternative methods to estimate $\L(\tb)$ marginals, e.g. \citet{Jeffrey2020} proposed using moment networks, which would be much cheaper than MCMC.} 
The MCMC method involves simulating a Markov chain in the parameter space, then the samples satisfy the target probability distribution function (PDF). Although the MCMC sampling method do not suffer from the computational complexity of multidimensional integration, it is still computationally expensive \citep{Auld2008}.  In addition, a representation of PDF with samples is noisy, and is inconvenient for further computations if an explicit expression of it like Eq. (\ref{like}) is unknown
\citep{Papamakarios2016}. 
More importantly, the method of MCMC based on $\chi^2$ analysis relies on the analytical specification of the likelihood (%
e.g. Eq. (\ref{like})), which is not often possible for problems with intractable likelihoods. 


To overcome the aforementioned limitations, we propose a procedure that includes a neural network model called Masked 
Autoregressive Flow (MAF) to constrain cosmological parameters. The model was initially proposed by \citet{Papamakarios2017} and is described in detail in Section \ref{sec:MAF}. 


\subsection{MAF for Parameter Constraint}
\label{sec:MAF}

\subsubsection{Masked Autoregressive Flows (MAF)}
\label{sec:MAF_struc}

Constraining cosmological parameters is essentially estimating the density distribution of $\tb$, usually the posterior $P(\tb|\Hb\obs)$. The density can be directly estimated using neural density estimators rather than sampling from the distribution with MCMC, provided that data $\Hb$ can be simulated with a simulation model given any possible parameter $\tb$.
Common density estimators include autoregressive models (e.g. Neural Autoregressive Distribution Estimation (NADE) proposed by \citet{Uria2016}, Masked Autoencoder for Distribution Estimation (MADE) proposed by \citet{Germain2015}, Inverse Autoregressive Flow (IAF) proposed by \citet{Kingma2015}), normalizing flows \citep{Com2015} and Real NVP \citep{Dinh2017}, among which MAF  achieved state-of-the-art performance on several examples described in \citet{Papamakarios2017}.

The 
design
of MAF is a combination of
autoregressive models and normalizing flows. An autoregressive model decomposes the joint distribution of a random vector, say $\tb$, into the product of its nested conditionals according to the chain rule of probability:
\begin{equation}
P(\tb) = \prod_{d=1}^{D} P(\theta_d|\tb_{1:d-1}),
\end{equation}
where $\tb_{1:d-1}=(\theta_1, \theta_2, ..., \theta_{d-1})\T$, $D$ is the dimension of the parameter vector.
Then it models each conditional $P(\theta_d|\tb_{1:d-1})$ as a parametric PDF, the parameters of which are functions of $\tb_{1:d-1}$. 
When modeling a conditional distribution $P(\tb|\Hb) = \prod_{d=1}^{D} P(\theta_d|\tb_{1:d-1}, \Hb)$, the parameters are functions of $(\tb_{1:d-1}, \Hb)$. For instance, the Gaussian MADE 
parameterizes $P(\theta_d|\tb_{1:d-1},\Hb)$ as Gaussian distributions with mean $\mu_d=
f_{\mu_d}(\tb_{1:d-1}, \Hb)$ and log standard deviation $\alpha_{d} = 
f_{\alpha_{d}}(\tb_{1:d-1}, \Hb)$, where $f_{\mu_d}, f_{\alpha_{d}}$ are free functions. In neural density estimation with autoregressive models, $f_{\mu_d}, f_{\alpha_{d}}$ are typically implemented with a neural network that inputs $(\tb_{1:d-1}, \Hb)$ and outputs parameters such as $\mu_d, \alpha_d$. 

It is unnecessary to model each conditional with a separate network, 
as a MADE \citep{Germain2015} can simultaneously output the parameters of all of these conditional distributions with one single network.
To construct a MADE that infer $D$ parameters from $N$-dimensional data (e.g. $\Hb\obs$), one should first construct a fully-connected multi-layer perception with $D+N$ inputs and $D$ outputs, and drop certain connections between the neurons to ensure that the parameters for the $d$-th conditional, outputted by the network, are only functions of $(\tb_{1:d-1}, \Hb)$ and are independent of $\theta_d, \theta_{d+1}, \cdots, \theta_D$. 
If a network satisfies this property---the so-called autoregressive property---the outputs can be interpreted as the estimation 
of conditional distributions $P(\theta_d|\tb_{1:d-1},\Hb)$.
The structure of MADE, satisfying the autoregressive property, is illustrated in Fig. \ref{fig:schemetictemp}. Practically, the connection dropping is actually done by multiplying each connection weight of the network with a binary "mask" (i.e. a number in $\{0,1\}$). See \citet{Germain2015} for more detail on MADE.

As pointed out in \citet{Papamakarios2017}, an autoregressive model, e.g. MADE, whose conditionals $P(\theta_d|\tb_{1:d-1},\Hb)$ are parameterized as Gaussian distributions, can be regarded as a normalizing flow. A normalizing flow \citep{Com2015} represents $P(\tb)$ with a base density $\pi(\bm{u})$ and an invertible transformation $f: \bm{u}\mapsto\tb=f(\bm{u})$. $\pi(\bm{u})$ is a simple distribution that can be easily evaluated, and is usually chosen to be a standard Gaussian ($\bm{u} \sim \mathcal{N}(\mathbf{0}, \bm{I})$), then density $P(\tb)$ can be evaluated by
\begin{equation}
P(\tb) = \pi(f^{-1}(\tb))\left|\det \left(\pp{f^{-1}(\tb)}{\tb}\right)\right|,
\label{eq:eval_density}
\end{equation}
where $\bm{u}=f^{-1}(\tb)$. For a Gaussian MADE, each of the conditionals $P(\theta_d|\tb_{1:d-1},\Hb)$ can be regarded as transformed from a standard Gaussian base density $\pi(u_d)$, i.e. 
\begin{equation} 
u_d\mapsto\theta_d=f_d(u_d; \alpha_d, \mu_d) = \exp(\alpha_{d})\cdot u_d+\mu_d,
\end{equation}
such that
\begin{gather}
\theta_d \sim P(\theta_d|\tb_{1:d-1},\Hb)=\mathcal{N}(\theta_d; \mu_d, \left(\exp \alpha_{d}\right)^{2}),\\
u_d\sim \pi(u_d) = \mathcal{N}(u_d;0,1),
\end{gather}
where
\begin{equation}
\mu_d=
f_{\mu_d}(\tb_{1:d-1}, \Hb), \quad \alpha_{d} = 
f_{\alpha_{d}}(\tb_{1:d-1}, \Hb).
\end{equation}
Thus, it can be proved that a Gaussian MADE is a normalizing flow with $\tb = f(\bm{u}; \Hb)$ where $\bm{u}\sim\mathcal{N}(\bm{0}, \bm{I})$.

The nature of normalizing flow makes it not only possible to evaluate the density (Eq. \ref{eq:eval_density}), but also feasible to sample from the learned distribution. To generate samples of $\tb$ from $P(\tb)$, one first generates $\bm{u}$  from the standard Gaussian distribution, then transforms it to $\tb = f(\bm{u})$. Inversely, a well-fitted normalizing flow can transform training data $\tb$ back to corresponding random numbers $\bm{u}=f^{-1}(\tb; \Hb)$ that approximately satisfy the standard Gaussian distribution, whereas a poor fitted normalizing flow outputs $\bm{u}$ that are not standard Gaussian.

The normalizing flow nature of MADE makes it possible to improve the performance by stacking several MADEs. A single MADE with Gaussian conditionals may not fit the distribution well if the real conditionals cannot be approximated by the Gaussian distribution. In addition, whether the real conditionals are similar to Gaussian may depend on the order of the components of $\tb$ (see \citet{Papamakarios2017} for discussion and examples).
If a MADE fails to fit the train data $\tb$ well, the corresponding random numbers $\bm{u}_1=f_1^{-1}(\tb; \Hb)$ are not Gaussian. \citet{Papamakarios2017} proposed to fit $\bm{u}_1$ with 
a second MADE, i.e. $\bm{u}_2 = f_2^{-1}(\bm{u}_1; \Hb)$, and $\bm{u}_2$ is fitted by the third MADE, etc. Thus, 
\begin{equation}
\tb = (f_1\circ f_2\circ\cdots \circ f_{N_{\mathrm{MADE}}})(\bm{u}_{N_{\mathrm{MADE}}}; \Hb),
\end{equation}
where $f_i$ is the invertible function corresponding to the $i$-th MADE, $N_{\mathrm{MADE}}$ is the number of stacked MADEs, $\bm{u}_{N_{\mathrm{MADE}}} \sim \mathcal{N}(\bm{0}, \bm{I})$. $f = f_1\circ\cdots \circ f_{N_{\mathrm{MADE}}}$ 
corresponds to
a new normalizing flow, which is called Masked Autoregressive Flow (MAF).


A MAF can be easily trained by minimizing the negative log probability (which is the loss function),
\begin{equation}
L = - \sum_n \ln P(\tb_n|\Hb_n),
\end{equation}
obtained from the MAF using the training set, $\{\tb_n,\Hb_n\}$.
In a series of experiments carried out by \citet{Papamakarios2017}, MAF outperformed Real NVP in all tasks and was better than a kind of MADE (whose conditionals are mixture of Gaussians) in about half of the tasks, showing the competitiveness of MAF.

\subsubsection{Likelihood-free Inference with MAF}
\label{sec:MAF_infer}
With a neural density estimator like the MAF, the likelihood-free inference can be turned into the estimation of the distribution of simulated parameter--data pairs, $\{\tb_n, \Hb_n\}$,
where parameters $\tb_n$ are sampled from a distribution $q(\tb)$, and data $\Hb_n$ corresponding to $\tb_n$ are simulated using the simulation model (specified in Eq. (\ref{like}) for OHD).
With regard to the specific distribution fitted by the model, there are three options,
as summarized and discussed in \citet{Alsing2019}: (1) the joint distribution $P(\tb, \Hb)$ \citep{Alsing2018}, (2) the conditional distribution\footnote{This is used as example in Section \ref{sec:MAF_struc}.} $P(\tb|\Hb)$ \citep{Papamakarios2016, lueckmann2017}, or (3) the conditional distribution\footnote{This is the distribution of uncertainty, which is Eq. (\ref{like}) for OHD.} $P(\Hb|\tb)$
\citep{Papamakarios2019, lueckmann2018}. With observed data $\Hb\obs$, the three options give estimations of the joint distribution $P(\tb, \Hb\obs)$, posterior $P(\tb|\Hb\obs)$ and likelihood $P(\Hb\obs|\tb)$ respectively. Thus we refer the options as (1) joint density estimation, (2) posterior estimation and (3) likelihood estimation.

Each option has advantages and disadvantages. First, it should be noted that the posterior density (which is a distribution of $\tb$) cannot be directly obtained using joint density estimation (fitting $P(\tb, \Hb)$) or likelihood estimation (fitting $P(\Hb|\tb)$), because the models can only output either the joint distribution or the distribution of $\Hb$ given the value of $\tb$. Thus, 
the posterior is obtained indirectly by
\begin{equation} 
P(\tb|\Hb\obs)\propto P(\tb, \Hb\obs)\propto P(\tb)P(\Hb\obs|\tb),
\end{equation}
where $P(\tb)$ is the prior distribution.
Since the value of posterior $P(\tb|\Hb\obs)$ can be obtained at arbitrary $\tb$, the distribution
can be 
estimated with traditional methods such as the MCMC. Conversely, posterior estimation (fitting $P(\tb|\Hb)$) directly outputs the posterior $P(\tb|\Hb\obs)$ given input $\Hb\obs$, which may save time of the second step of estimation. It is also straightforward to sample from the posterior using posterior estimation.

Second, with regards to joint density estimation or posterior estimation, since the model directly models  the distribution of $\tb$ (or the joint distribution of $\tb, \Hb$), the $\tb$ for simulated $\{\tb, \Hb\}$ are sampled from the whole prior $P(\tb)$. This may result in the need for more simulations of $\{\tb, \Hb\}$ to get an accurate estimation. Alternatively, $\tb$ can also be sampled from a proposal distribution $q(\tb)$, and posterior should be re-weighed by $P(\tb|\Hb\obs)/q(\tb)$, but this makes it indirect to evaluate or generate samples from the posterior. On the other hand, for likelihood estimation, the distribution of $\tb$ for simulated training data, $q(\tb)$, does not affect the resulting posterior. The only effect is that $P(\Hb|\tb)$ is modeled more accurately where the density $q(\tb)$ is larger. Given this fact, a sequential method can be adopted to make $P(\Hb|\tb)$ more accurate near more likely parameters (see \citet{Papamakarios2019} for details).

Thus, we suggest that posterior estimation (fitting $P(\tb|\Hb)$) can be used when it is inexpensive to simulate data,
the prior is not much larger than the posterior (this is discussed below), or the value of the posterior distribution is used for subsequent computations. On the other hand, likelihood estimation (fitting $P(\Hb|\tb)$) can be used when simulation is very expensive (so the number of simulation is limited), or it is inconvenient to generate train data $\{\tb, \Hb\}$ with $\tb$ satisfying the prior $P(\tb)$.

In our experiments, we choose to use posterior estimation, directly fitting the conditional density $P(\tb|\Hb)$, for the convenience of directly evaluating and sampling from the posterior. 



\begin{figure*}[ht]
\centering
\includegraphics[width=0.73\linewidth]{%
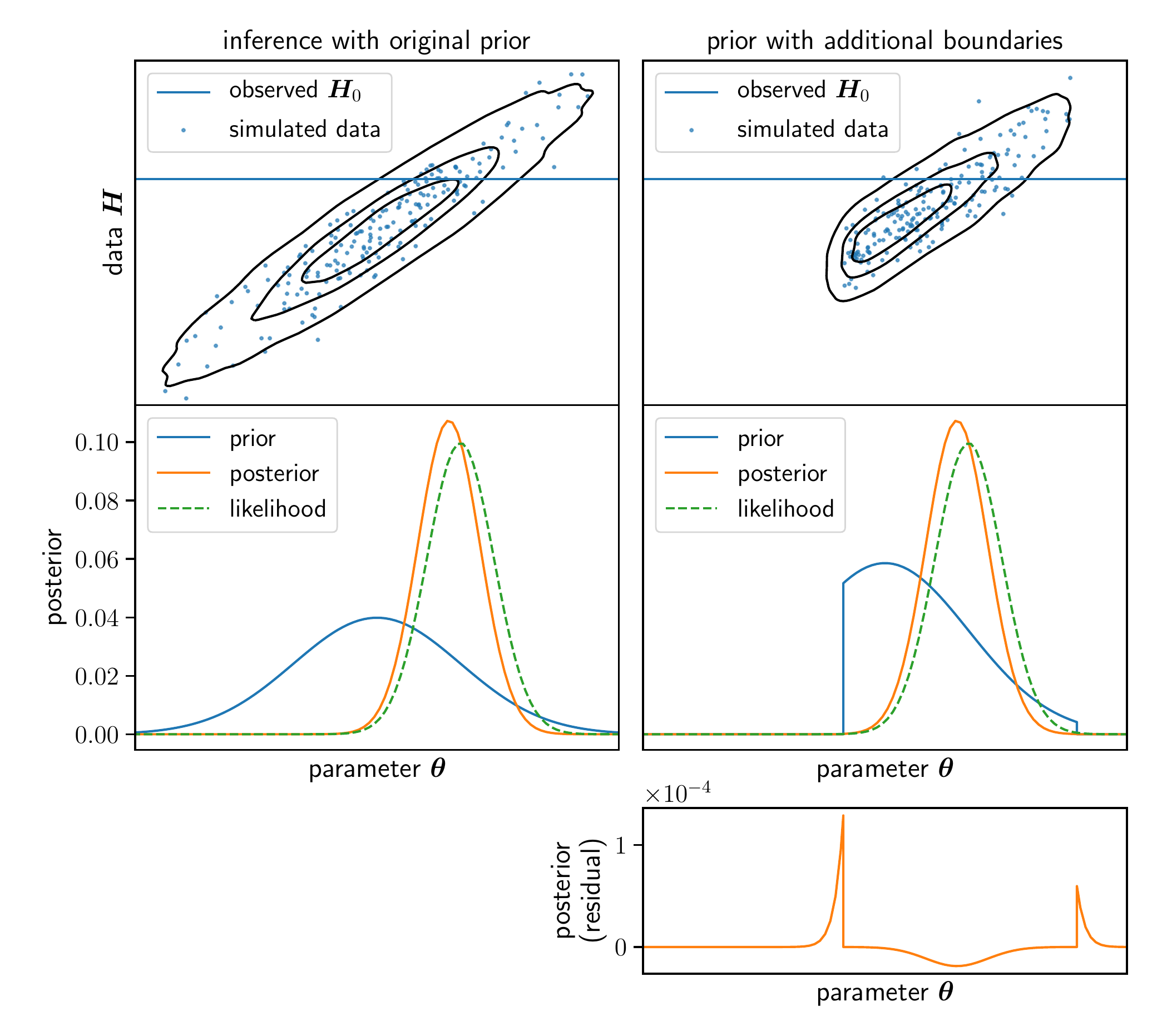}
\caption{Artificially adding boundaries to the prior $P(\tb)$ has no practical effect on the posterior $P(\tb|\Hb\obs)$, provided that the boundaries encloses the likely region of the posterior. (1) The left panel: parameters $\tb$ are sampled from a broad prior, and corresponding data $\Hb$ are simulated to get the joint distribution of $\{\Hb, \tb\}$. The posterior $P(\tb|\Hb\obs)$ is the conditional density at observed data $\Hb=\Hb\obs$, and is learned by MAF in this paper. Only samples near $\Hb=\Hb\obs$ contribute to the $P(\tb|\Hb\obs)$ learned by MAF. (2) The right panel: the samples of $\tb$ are restricted to a range, which means adding additional boundaries to the prior. The posterior $P(\tb|\Hb\obs)$ is practically equal to that of the left panel, but less simulation is needed to achieve the same accuracy. The difference between the two posteriors (with and without additional boundaries) is small, as shown in the residual plot on the bottom right. Thus, it may be unnecessary to simulate training data in the whole prior.}
\label{fig:prior_prob}
\end{figure*}

When fitting $P(\tb|\Hb)$, one issue that may affect the performance of the MAF is the selection of the prior. 
First, when we assume a prior $P(\tb)$ that is much broader than the 
likely
region (e.g. the 99.74\% confidence region) of the posterior (as illustrated in the left panel of Fig. \ref{fig:prior_prob}), much simulations are generated at $\tb$ where the likelihood is close to zero. Unfortunately, these simulations have no significant effect on the evaluation of the posterior $P(\tb|\Hb\obs)$, 
and
more simulations are needed to guarantee enough data within the likely region of the posterior.  
Secondly, when the prior $P(\tb)$ has hard boundaries  (where the density abruptly drops to 0), the posterior may also have hard boundaries. However, it may need much more MADEs to transform from a standard Gaussian base density to a density with hard boundaries, and more MADEs may require more training data. 
Therefore, to obtain a reasonable inference with less simulation, one may avoid learning posterior distributions with hard boundaries and restrain simulation near likely regions of the posterior. Unfortunately, the prior distributions in real problems are determined by our prior knowledge and should not be artificially changed.

We propose that one way to alleviate the aforementioned problems is to sample $\tb$ from a distribution $q(\tb)$ similar to the prior $P(\tb)$ but has different
hard boundaries. For convenience, we define $P'(\tb)$ as the prior extended to the whole parameter space, such that it is proportional to the prior $P(\tb)$ within its hard boundaries.\footnote{If $P(\tb)$ has no boundaries, $P'(\tb)=P(\tb)$; they can be considered to have infinite boundaries.} Then, we can choose $q(\tb)$ to be $P'(\tb)$ with additional hard boundaries
that sufficiently enclose the
likely
region
of
 $P'(\tb)P(\Hb\obs|\tb)$ 
but are not too broad to enclose the extremely unlikely region of $P'(\tb)P(\Hb\obs|\tb)$. As illustrated in Fig. \ref{fig:prior_prob}, when hard boundaries that enclose the likely region of the posterior are added to the prior, the new posterior is almost the same as the original one. If the chosen boundaries of $q(\tb)$ are broader than those of the actual prior $P(\tb)$, the posterior can be simply evaluated by generating samples from MAF and discarding those outside the boundaries of $P(\tb)$. If the opposite is true, although $\tb$ are not generated in the whole possible region of the prior $P(\tb)$, the distribution learned by a MAF is hopefully still a good approximation of the posterior (as shown in Fig. \ref{fig:prior_prob}). 

It may be tricky to choose the boundaries for $q(\tb)$ when there is little knowledge about where the likely region of $P'(\tb)P(\Hb\obs|\tb)$ might be in the parameter space. In this case, we propose a sequential procedure%
, inspired by \citet{Papamakarios2019, Papamakarios2016, lueckmann2017flexible}.
We propose that 
the MAF can be re-trained several times: in each round, the MAF is trained with $\tb$ sampled from a new distribution $q(\tb)$ whose boundaries sufficiently enclose all the samples generated from the distribution $P(\tb|\Hb\obs)$ estimated in the last round. In the first round, the boundaries are chosen to
enclose all possible regions, so in several rounds the region will be smaller and better match the likely region of the posterior.
Additionally, another way is to assume an approximate analytical form of the likelihood, estimate the approximate posterior with MCMC, and choose boundaries of $q(\tb)$ that sufficiently enclose the samples from MCMC.\footnote{Note that the tails of an MCMC are the least likely to be in the chain, and one may need to choose broader boundaries to avoid underestimation of the likely region.}


For the sake of simplicity, we use uniform priors $P(\tb)$ in the experiments to demonstrate and evaluate the sequential procedure proposed above. With uniform priors, the distributions $q(\tb)$ are also uniform, and the ranges of $q(\tb)$ are chosen in the sequential procedure.

Finally, it is worth noting that although the uniform prior set in this paper is often used in similar works \citep[e.g.][]{Weyant2013}, the real non-informative prior is Jeffreys prior rather than uniform prior. According to Jeffreys' rule, the non-informative prior is taken to be proportional to the square root of the determinant of the Fisher information matrix $I^F$ \citep[see e.g. ][]{bayesianinference}, i.e. 
\begin{equation}
P_{\mathrm{Jeffreys}}(\tb) = \frac{\sqrt{\det\left[I^F(\tb)\right]}}{\int\mathrm{d}\tb \sqrt{\det\left[I^F(\tb)\right]}},
\end{equation}	
where the Fisher information matrix as a function of parameters $\tb$ is defined as 
\begin{equation}
I^F_{\alpha\beta}(\tb) = 
\int\mathrm{d}\Hb P(\Hb|\tb)
\left\{-\frac{\partial^2}{\partial\theta_\alpha\partial\theta_\beta} \ln P(\Hb|\tb)\right\}.
\end{equation}
However, for intractable likelihoods dealt with in this paper, one can only obtain an numerical estimation of $I^F(\tb)$ with likelihood estimated using MAF, and calculating or sampling from the prior involves high-dimensional numerical integration, which is computationally expensive. 


\subsection{Autoencoder for 
	Data 
	Dimensionality Reduction
}\label{sec:ae}

It is worth noting that a MAF is still a parametric model, although it is implemented with a neural network that has thousands or even millions of parameters. Therefore, a MAF can only express transformations in a subset of function space and approximate the target density. As mentioned in Section \ref{sec:MAF_struc}, a MAF improves the performance of a Gaussian MADE by stacking them. When the posterior distribution is irregular, many MADEs may have to be used to transform the standard Gaussian base density to the posterior. However, for a problem where the dimension of data is very large, training a large MAF that consists of several cumbersome MADEs is too expensive. In addition, more complex networks require larger datasets to train, but generating a large number of data using the simulation model is usually too time consuming. 

Thus, rather than the
high-dimensional data itself, the statistics of data or some empirical parameters were used to 
perform inference in e.g. \citet{Papamakarios2016}.
We propose that
data $\Hb$ can be represented by lower dimensional encodings 
learned with another kind of commonly-used neural network, namely denoising autoencoder (DAE). Since encodings are automatically learned with DAE, data can be better represented without any artificial choices of statistics.


Autoencoders are capable of learning latent representations of data \citep{vincent2008}. A basic autoencoder is composed of an encoder and a decoder. The encoder learns to map input data $\Hb$ to lower-dimensional encodings (or feature, latent representation) $\bm{y}=f_e(\Hb)$, while the decoder learns to do the opposite -- reconstruct the input data of encoder given the encodings, $\Hb' = f_d(\bm{y})$.
The parameters of the autoencoder are usually optimized by minimizing reconstruction error, i.e. the mean squared error (MSE) between reconstructed data $\Hb'$ and the label. 
Though the label is usually the real input data of the 
encoder,
what represents most of the information of $\Hb$ is the noise-free part rather than the Gaussian noise. Thus, for OHD we used 
noise-free fiducial values $\Hb\fid$
as labels, while the inputs were noisy data, so that an autoencoder inputs noisy data and tries to output data that are close to the noise-free data. This makes the model a stacked denoising  autoencoder (DAE, \citet{dae}), which can not only learn the robust features but also significantly reduce the noise level. 
	
Representing high-dimensional data $\Hb$ with encodings $\yb$, we can use MAF to learn the posterior $P(\tb|\yb)$ instead of $P(\tb|\Hb)$, where $\yb=f_e(\Hb)$ contains much of the information in $\Hb$. Since the dimension of $\yb$ is smaller than $\Hb$, a smaller MAF (with less hidden layers and less neurons) can be used to get a reasonable posterior. A smaller MAF needs less data to converge, making it more computationally efficient. Although some simulations may be needed to train the DAE,  it is worthwhile to make the inference step using MAF faster and less resource consuming. This is because for high-dimensional data, a MAF with several layers of MADEs is usually larger than a DAE, and thus consumes most of the computational resources.


In this paper, we train the DAE to not only minimize the reconstruction MSE, but also get encodings $\yb$ such that the variance of $P(\yb|\tb)$ for given $\tb$ is as small as possible.
This pushes the posterior $P(\tb|\yb)\propto P(\tb)P(\yb|\tb)$ to have smaller variance, and thus 
the DAE will try to avoid outputting $\yb$ that gives too big posteriors of $\tb$, and $\yb$ may contain more information about $\tb$. To make this possible, we first require that the mean of $\yb$ given $\tb$, i.e. the mean of the conditional $P(\yb|\tb)$, relies linearly on $\tb$. Noting that $f_e, f_d$ are non-linear transformations implemented with neural networks, the linear relationship does not necessarily hold for $P(\Hb|\tb)$.

With that requirement, an additional loss function of the autoencoder in addition to the reconstruction error can be constructed. First, a linear regression model is fitted for a batch of training data, i.e. the linear prediction $\yb_{\mathrm{pred}}$ is given by
\begin{equation}
\yb_{\mathrm{pred}} = (1, \tb\T)\bm{A},
\end{equation} 
where $\bm{A}$ is the matrix of linear coefficients. $\bm{A}$ can be easily calculated with $\bm{A} = \vT^{+}\bm{Y}$, where
\begin{equation}
\bm{Y}=
\begin{pmatrix}
\bm{y}_1\T\\
\bm{y}_2\T\\
\vdots
\end{pmatrix},\quad
\vT = 
\begin{pmatrix}
1&\tb_1\T\\
1&\tb_2\T\\
\vdots&\vdots
\end{pmatrix}
,
\end{equation}
and $\vT^+$ is the pseudoinverse (Moore-Penrose inverse) of $\vT$. 
Then, the variance of $\yb-\yb_{\mathrm{pred}}$ is calculated and used as the additional loss function. This additional term not only penalizes the model when the relationship between $\yb$ and $\tb$ is not linear, but also pushes the encoder to output $\yb$ with smaller variances.

Therefore, the complete batch loss proposed in this paper for autoencoder 
consists of reconstruction MSE and encoding variance: 
\begin{equation}
\begin{aligned}
L_{\mathrm{AE}} =&
\operatorname{mean}\left\{\left(\Xb'-\Xb_{\mathrm{fid}}\right)\circ \left(\Xb'-\Xb_{\mathrm{fid}}\right)\right\}\\
&+ \operatorname{var}\left\{\bm{Y} - \vT\vT^{+}\bm{Y}\right\}
\end{aligned}
\end{equation}
where 
\begin{equation}
\X_{\mathrm{fid}} =
\begin{pmatrix}
\Hb_{\mathrm{fid}, 1}\T\\
\Hb_{\mathrm{fid}, 2}\T\\
\vdots
\end{pmatrix},\quad
\X' =
\begin{pmatrix}
{\Hb'_1}\T\\
{\Hb'_2}\T\\
\vdots
\end{pmatrix},
\end{equation}
$\Xb_1\circ \Xb_2$ denotes the Hadarmard (element-wise) product of matrix $\Xb_1$ and $\Xb_2$, $\operatorname{mean}\left\{\Xb\right\}$ and $\operatorname{var}\left\{\Xb\right\}$ denote the mean and variance value of the elements of $\Xb$. The loss defined as above can be easily evaluated on the training set.



To train an autoencoder constructed as above, one usually needs to generate 
training data using the simulation model.
For OHD discussed in this paper, one should first assume a cosmological model,
and calculate sufficient instances of noise-free $\Hb$ corresponding to a variety of arbitrary parameters $\tb$ of the model. Unlike MAF, the $\tb$ for training data for DAE can be sampled in any range, and DAE gives reliable dimensionality reduction if the most likely values of $\tb$ are in the range. During training, Gaussian noise with mean 0 and the standard deviation $\sb$ of the real OHD $\Hb\obs$ is added 
before inputted to the DAE.
\footnote{When using data like the OHD, there is an alternative way to construct and train the autoencoder: during training, the noise is not added outside the autoencoder, but rather added by an additional noise layer inside the autoencoder. The noise layer is shut down when using the encoder to encode the real OHD. This kind of network with noise layers is commonly used in machine learning.}

\

\begin{figure*}
	\centering
\includegraphics[width=0.95\linewidth]
{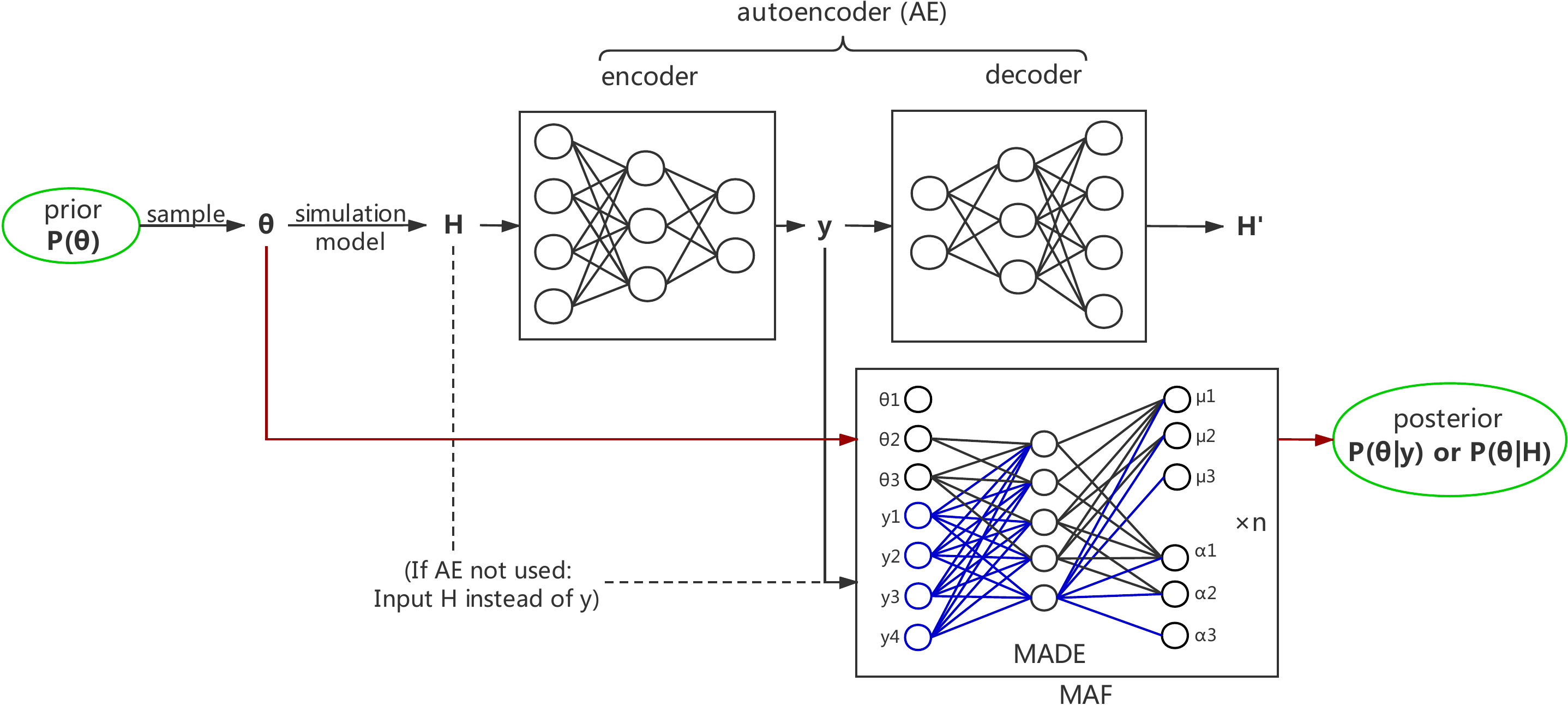}
	\caption{An illustration of the structure of the models and the procedure proposed in this work.
	The upper half is a demonstration of autoencoder, while the lower part shows the MAF. During the training of the autoencoder, simulated data $\Hb$ (with noise) corresponding to different $\tb$ is generated, and the autoencoder is trained to encode $\Hb$ to encodings $\bm{y}$ and decode $\bm{y}$ to output reconstruction result $\Hb'$ that is close to noise-free data $\Hb$. During training of MAF, $\tb$ is sampled from prior, and $\Hb$ corresponding to $\tb$ is simulated from the 
	simulation model,
then MAF is trained on $\Hb$ (or the encoding $\bm{y}$ if autoencoder is used) and $\tb$. When evaluating cosmological parameters, MAF evaluates the posterior function \replaced{$P(\tb)$}{$P(\tb|\Hb\obs)$ (or $P(\tb|\yb\obs)$)} in parameter space given observed Hubble parameter $\Hb\obs$ (or $\bm{y}\obs$). }
	\label{fig:schemetictemp}
\end{figure*}

The whole procedure of constraining the parameters with DAE and MAF can be summarized as below:
(1) Generating 
data with the simulation model
and training a DAE on it;
(2) Generating 
training data from the simulation model and encoding the data with the encoder of the DAE to get lower-dimensional encodings;
(3) Training a MAF on the encodings and corresponding parameters
with the proposed sequential process;
(4) Encoding the real OHD with the encoder of the DAE and inputting
{the encodings} 
to the MAF to evaluate the posterior of parameters. The structure of the models and the procedure are shown in Fig. \ref{fig:schemetictemp}.

\section{Evaluation on Simulated OHD}
\label{sec:simul}
\subsection{The Simulated Data}
To evaluate the ability of the proposed procedure to constrain parameters in cosmology, the ANNs were first applied to mock OHD data. During the evaluation in this section, the mock OHD data is regarded as the "real" data, and the ANNs are trained on the training set $\{\tb, \Hb\}$ to infer the parameters of the mock OHD. Both mock OHD and ANN training data were generated using the \La CDM simulation model described below.

Our simulation model is based on the 
non-flat \La CDM model and the simple multivariate Gaussian uncertainty in Eq. (\ref{like}) and (\ref{chi2}), so that ANNs can be compared with traditional methods. The fiducial value of $H(z)$, denoted as $H\fid(z)$, is given by Eq. (\ref{eq:fiducial}), with 3 free parameters $H_0, \Om, \Ol$. To test the performance of our method near the most probable parameters, we always generated the mock OHD 
with inferred parameter values from Planck observations:
$H_0=67.4\ \mathrm{km\ s^{-1}\ Mpc^{-1}}$, $\Om=0.314$, $\Ol=0.686$ \citep{Planck2013, Ia2018}, although parameter $\tb$ is not fixed when generating training data for ANNs.

\begin{figure}[tbh]
	\centering
	\includegraphics[width=.48\textwidth]{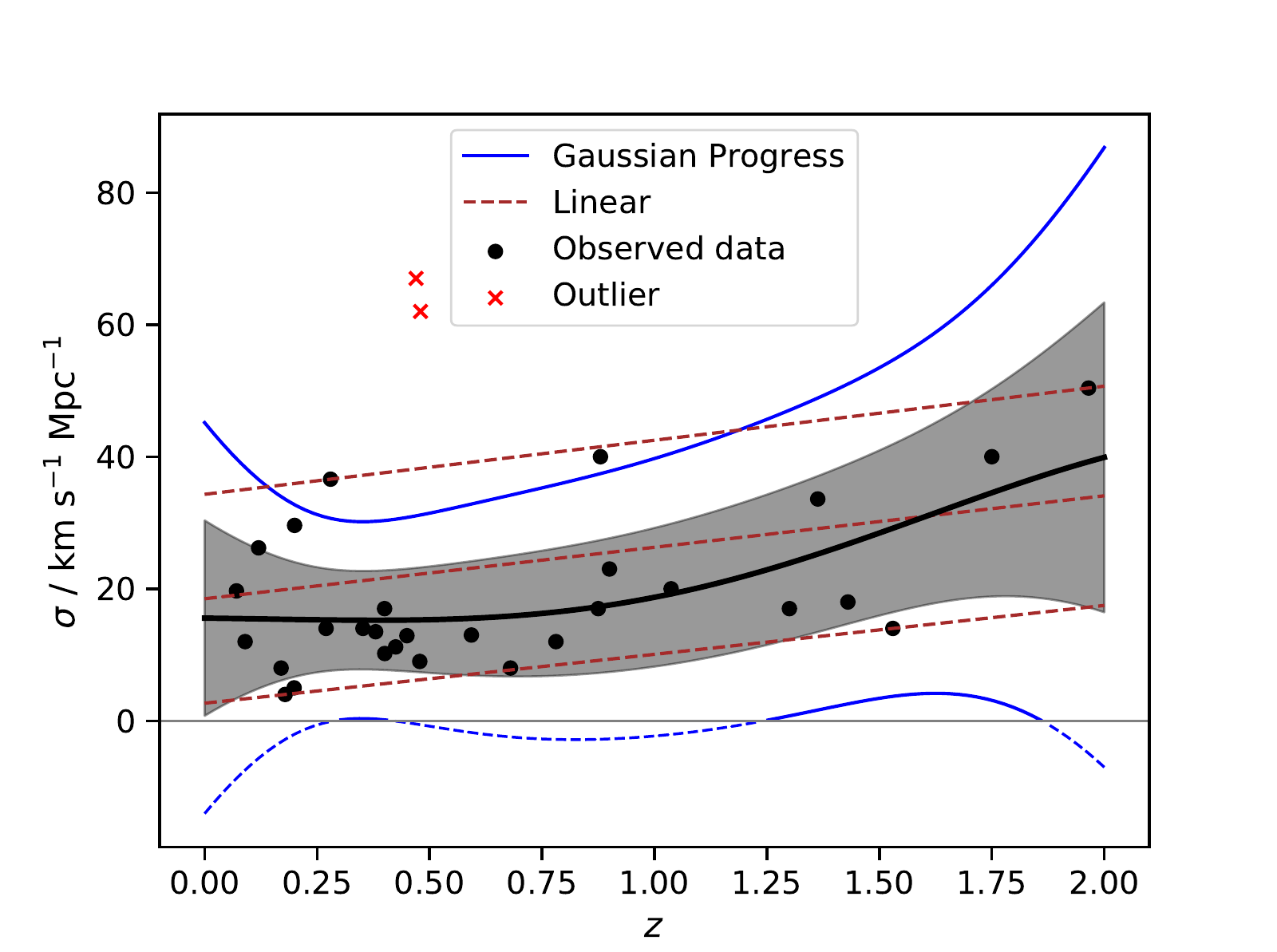}
\caption{Modeling the change of uncertainty $\sigma$ of $H(z)$ with redshift $z$ using Gaussian Process (solid curved lines). The  existing 31 real OHD data evaluated with cosmic choronometer method is marked as black dots (the two outliers are marked as red crosses). The mean, $1\sigma$ and 2$\sigma$
confidence region obtained with Gaussian process is represented as 
black thick solid line,
dark area and solid blue line respectively. We add a hard boundary, requiring $\sigma>0$ by discarding the values of $\sigma$ that are less than or equal to zero. The upper bound $\sigma_+(z)$, lower bound $\sigma_-(z)$ and mean $\bar{\sigma}(z)$ of the linear method in \cite{Ma2011} is plotted with red dashed lines.  
}
	\label{sigmaz}
\end{figure}

With respect to the relationship between uncertainty $\sigma$ and $z$, there is no well-defined simulation model for OHD due to the absence of future OHD survey plans. Therefore, such models are heuristically established by inspecting the uncertainties $\sigma(z)$ of existing 31 OHD points (introduced in Section \ref{sec:OHD}), as shown in Fig. \ref{sigmaz}. The general trend of existing $\sigma(z)$ is that the uncertainty increases with redshift, with 2 outliers excluded. The uncertainty models mentioned in references are either setting a relative error of $H$ \citep[e.g.][]{Wang2012,Moresco}, or bounding the uncertainties with two straight lines ($\sigma_+(z)$ and $\sigma_-(z)$) and sampling $\sigma(z)$ from a Gaussian distribution between the two lines \citep[e.g.][]{Ma2011,Seikel2012,Busti2014, Wang2020}. 
However, the former method ignores the uncertainty of $\sigma$ at given redshifts; the latter adopts a Gaussian distribution, but used mean $\bar{\sigma}(z)=[\sigma_+(z)+\sigma_-(z)]/2$ and standard variation $\tilde{\sigma}(z)=[\sigma_+(z)-\sigma_-(z)]/4$ that are linear functions of $z$, which do not perfectly model the trend of $\sigma(z)$.


In this work, to get a more realistic model of $\sigma(z)$, we propose  that the uncertainty $\sigma$ can be reconstructed with Gaussian process (with a hard boundary $\sigma > 0$). The observations of sigma $\bm{\sigma}=(\sigma_{ i})\T$ at each point $z_i$ are still assumed to satisfy the Gaussian distribution, i.e.
$\bm{\sigma}\sim\mathcal{N}(\bar{\bm{\sigma}}, \bm{K})$ where $\bar{{\sigma}}_i = \bar{{\sigma}}(z_i)$,  $\bar{\sigma}$ is the mean function, $(\bm{K}_{ij}) = (\operatorname{cov}\left(\sigma_{i}, \sigma_{j}\right))$ is the covariance matrix.
However, instead of empirically setting linear mean and standard deviation, Gaussian process optimizes the mean and covariance 
to maximize the likelihood $P(\sb\obs|\bar{\bm{\sigma}}, \bm{K})$.
The covariance matrix is given by 
\begin{equation}
\operatorname{cov}\left(\sigma_{i}, \sigma_{j}\right)=k\left({z}_{i}, {z}_{j}\right)+\tilde{\sigma}_{i}^{2} \delta_{ij},
\end{equation}
where $\delta_{ij}$ is Kronecker delta, $k$ is kernel function (or covariance function) \citep{Ia2018, GPbook}, and $\tilde{\sigma}_i$ indicates the noise level of sigma at $z_i$ (the uncertainty of $\sigma$). 
We adopt the commonly used kernel function for this type of work, namely the Gaussian radial basis function (or squared exponential function):
\begin{equation}
k\left(z_{i}, z_{j}\right)=\sigma_{f}^{2} \exp \left(-\frac{\left(z_{i}-z_{j}\right)^{2}}{2 l^{2}}\right),
\end{equation}
where $\sigma_f$ and $l$ are hyperparameters, and are optimized during training. The values $\tilde{\sigma}_i$ (added to the diagonal of $\bm{K}$) are assumed to be the same at each point in this work:
\begin{equation*}
\tilde{\sigma}_i=\tilde{\sigma},\quad i=1,2,...
\end{equation*}
and $\tilde{\sigma}$ is optimized with grid search with a 4-fold cross validation. The above process was implemented with \lstinline
|scikit-learn| \citep{scikit-learn}. 

The Gaussian process reconstruction of $\sigma(z)$ is shown in Fig. \ref{sigmaz}. As can be seen, the trend of existing OHD $\sigma$ is better described with a curve than a straight line. To make a more quantitative comparison, we repeated the linear method in e.g. \citet{Ma2011} to bound the data with two straight lines, and set the linear mean and standard variation, as shown in Fig. \ref{sigmaz}. We found that the log likelihood $\ln P(\bm{\sigma}\obs|\bar{\sb}, \bm{K})$ is $-105.97$ for the Gaussian process and $-113.94$ for the linear method, showing that the Gaussian process method better models the relationship between $\sigma$ and $z$.

To simulate OHD, we first draw $\z=(z_i)\T$
from a uniform distribution in $[0,2]$; then the corresponding $\sig=(\sigma_i)\T$ is drawn according to the result of the Gaussian process. 
Noting that another constraint on the uncertainties is $\sigma_i>0$, negative samples of $\sig$ are discarded. Finally, the simulated $\Hb\obs = (H_{\mathrm{obs},i})\T$ is sampled according to $H_{\mathrm{obs},i}=H\fid(z_i)+\Delta H_i$, where $\Delta H_i \sim\N(0,\sigma_i^2)$.

It is still worth noting that the Gaussian process used in this paper is still an empirical model, estimating the uncertainties by fitting the existing data.
Though this does not invalidate the comparison between our procedure and the traditional method, this simulation model cannot describe the physical mechanism of $H(z)$ data generation. 
Future works should focus on better models to simulate OHD, considering the original procedures of measuring $H(z)$ from observational data, e.g. galaxy spectrum (the cosmic chronometer method proposed by \citet{Jimenez2002}).

\subsection{Criteria for Posterior Comparison}
\label{sec:goodness}

To evaluate the performance of the procedure in this paper, one straightforward approach is to compare the posterior learned with MAF to that directly calculated using the analytical expression Eq. (\ref{like}), but directly calculating the real posterior is too computationally expensive. Alternatively, the posterior of MAF is compared with that of the standard MCMC method using the KL divergence (described later in this section). Apart from the standard KL divergence criterion, some other criteria that indicate the size and position of confidence regions may provide
a glimpse into the properties of the posteriors, and can be used to make comparisons. Thus, we also used 3 relative goodness of constraint
(figure of merit, log probability of true parameters and distance to the real parameter). 

In the following paragraphs, we will define the four criteria and discuss their meanings. For some of the criteria, the Planck \replaced{cosmological}{best fit} parameters\deleted{, which are used to generate mock $H(z)$ observations,} are used \replaced{to compare}{when comparing} the posteriors. \replaced{The Planck parameters are referred to as "true parameters", since these are the true values one is trying to recover by observing 
$H(z)$ and doing the Bayesian inference (though one can never get the exact true values by measuring).}{Because we generate the mock data at the Planck best fit parameters and try to infer their values, we call them true parameters, in the sense that we know the values used to generate the data. } 

\textit{Kullback-Leibler divergence (KL divergence)}. The KL divergence has a close relation to Fisher information and some other quantities of information theory, e.g. Shannon entropy, cross entropy, mutual information, etc. It is one of the common measures of the difference of probability distributions, $p_1(\tb)$ and $p_2(\tb)$, showing the information loss of approximating $p_1(\tb)$ using $p_2(\tb)$. Thus, they can be used to directly measure the difference between the posterior of MAF and that of the standard MCMC. The KL divergence from $p_1(\tb)$ to $p_2(\tb)$ is defined as 
\begin{equation}
D_{\mathrm{KL}}
\left(p_1(\tb) \middle\| p_2(\tb)\right) = \mathbb{E}_{p_1(\tb)} \left(\log p_1(\tb)-\log p_2(\tb)\right).
\end{equation}
It can be seen that one limitation of KL divergence is that it diverges at points outside the support of $p_2(\tb)$, i.e. $p_2=0$. 
Since the posterior of the MAF is never equal to zero, the above problem does not exist and KL divergence can be used.

In this paper, with $M$ samples $\{\tb_i\}$ from MCMC, the KL divergence from the posterior of MCMC $P_1(\tb|\Hb\obs)$ to that of MAF $P_2(\tb|\Hb\obs)$ (with and without DAE) is estimated with
\begin{equation}
D_{\mathrm{KL}}
\left(P_1 \middle\| P_2\right) = \frac{1}{M} \sum_{i=1}^M \left(\ln P_1(\tb_i|\Hb\obs)-\ln P_2(\tb_i|\Hb\obs)\right).
\end{equation}
For a neural density estimator like MAF, the probability $P(\tb|\Hb)$ at any point in the parameter space can be readily calculated given OHD data $\Hb$. For sampling methods like MCMC, this can be calculated with kernel density estimation (KDE).
The smaller the KL divergence, the closer the posterior of MAF is to that learned with MCMC.  Zero KL divergence means the equivalence of the posterior of MCMC and MAF.


\textit{Figure of merit}. In relevant works, a variety of figure of merits (FoM) were defined to quantify the ability of datasets and methods to tighten the constraints. In this work, a statistical FoM similar to the definition of \citet{Ma2011} and \citet{Wang2012} is adopted. The FoM is defined as the reciprocal volume of the confidence region of the posterior, which is given by a contour
\begin{equation}
P(\tb|\Hb\obs) = \text{const.} = \exp{(- \Delta\chi^2/2)}P_\ma,
\label{eq:chi2}
\end{equation}
where
$P_\ma$ is the maximum possibility density of the posterior, and $\Delta\chi^2$ is a constant. The constant $\Delta\chi^2$ is set so that $\exp{(- \Delta\chi^2/2)}P_\ma$ is equal to the probability density at the boundary of the $95.44\%$ confidence region of the Gaussian distribution.\footnote{Note that $\Delta\chi^2$ corresponds to $(\tb-\bm{\mu})\T\bm{K}^{-1}(\tb-\bm{\mu})$ for Gaussian distribution with mean $\bm{\mu}$ and covariance matrix $\bm{K}$, but can be regarded as a set constant because the posterior is usually non-Gaussian.} In this paper, $\Delta\chi^2$ is set to 8.02.
\footnote{The $\Delta\chi^2$ corresponding to the $95.44\%$, etc. confidence regions depend on the dimension of the space. In this paper, the free parameters consist of $H_0, \Om, \Ol$ and the parameter space is 3-dimensional. In this case, $\Delta\chi^2$ is approximately 8.02.}
Since FoM shows the volume of the confidence region, we can learn that the MAF cannot give tight constraint if its FoM is smaller than that of MCMC, or the MAF is misbehaving if it gives a constraint tighter than MCMC.

\textit{Log probability of true parameters $\ln P(\tb_\mathrm{true}|\Hb\obs)$.} A common way to quantify the accuracy of parameter constraints is to calculate the value of negative log probability of the posterior distribution at the true parameter point, as used in the experiments of \citet{\Papa2019}. A large probability of the true parameters suggests that the maximum of posterior is closer to true parameters or the confidence region is smaller, so one better recovers the true parameters with the Bayesian inference. Similar to the discussion for FoM, the MAF should recover true parameters as well as the MCMC given the same set of mock $H(z)$ observations.
When the training set for the MAF is not representative of the true generative process or the MAF fails to fit the data distribution well, there should be a difference between the values of $\ln P(\tb_\mathrm{true}|\Hb\obs)$ for MAF and MCMC.

\textit{Distance to the true parameters.} After estimating the maximum a posteriori (MAP) point, we calculate its Euclidean distance to the true parameters (abbreviated to $d$ hereafter) and use it as a criterion of the goodness of constraint. This criterion indicates to some extent the position of the confidence regions of posterior. For instance, when there is a great difference between the $d$ of MCMC and MAF, the posterior of MAF may be deviated from the expected position.


Unlike KL divergence, the last three criteria can be evaluated separately for each method, and the relative value between MAF and MCMC should be defined. In this work, the relative FoM and $d$ are defined as 
\begin{equation}
\begin{aligned}
\mathrm{FoM (relative)} &=\mathrm{ FoM (MAF) / FoM (MCMC)}, \\
d \mathrm{(relative)} &= d \mathrm{(MAF)} / d\mathrm{(MCMC)},
\end{aligned}
\end{equation}
while the relative $\ln P(\tb_\mathrm{true}|\Hb\obs)$ (abbreviated to $\ln P$ hereafter) are defined as 
\begin{equation}
\begin{aligned}
\ln P (\mathrm{relative})& = \ln P (\mathrm{MAF}) - \ln P (\mathrm{MCMC}).
\end{aligned}
\end{equation}
Thus, 1 (for FoM and $d$) or 0 (for $\ln P$) relative value means the equivalence of MAF's and MCMC's goodness of constraint. Since the posterior estimated using MCMC is the standard result, the closer these relative values are to 1 (or 0), the more similar the posterior of MAF is to that of MCMC, the better MAF performs.

Finally, it should be keep in mind that the above comparison of the MAF and the MCMC 
only shows
the performance of the MAF when the likelihood for the MCMC accurately describes the true distribution of the data. For the experiments in this section, the mock $H(z)$ data are generated according to the Gaussian likelihood Eq. (\ref{like}), so the posterior of the MCMC is the standard result. In most cases the likelihood for MCMC is only an approximation, so the result of MCMC can be biased, and a well-performed MAF is not necessarily similar to the MCMC.

\subsection{Experiments and Results}
\label{sec:SimRes}
In this paper, the neural networks described in Section \ref{sec:MAF} and \ref{sec:ae} are implemented  mainly with 
\lstinline|Keras| \citep{chollet2015keras}, \lstinline|TensorFlow| \citep{tensorflow} and an open source \lstinline|TensorFlow| implementation of MAF\footnote{This \lstinline|TensorFlow| implementation can be found at \url
{https://github.com/spinaotey/maf\_tf}, which is an adaptation from G. Papamakarios' \href{https://github.com/gpapamak/maf/}{maf}
 classes implemented with \lstinline|Theano|.}. 

\begin{table}
	\centering
	\caption{Number of neurons in hidden layers of each MADE of the MAF in the experiment}
	\label{tab:MAFconfiguration}
\begin{tabular}{c>{\centering\arraybackslash}p{2.2cm}>{\centering\arraybackslash}p{2cm}}
	\hline\hline
	Data dimension $N$ & Hidden layers (without DAE)&Hidden layers (with DAE) \\
	\hline
	300 &  120, 120, 120 & 20, 20\\
	200 &  80, 80, 80 &  20, 20 \\
	100 &  60, 60, 60  &  20, 20\\
	50 &  40, 40 &  20, 20\\
	30 & 20, 10  &   20, 20 \\
	\hline
\end{tabular}
\end{table}

\begin{figure}
	\centering
	\includegraphics[width=0.9\linewidth]{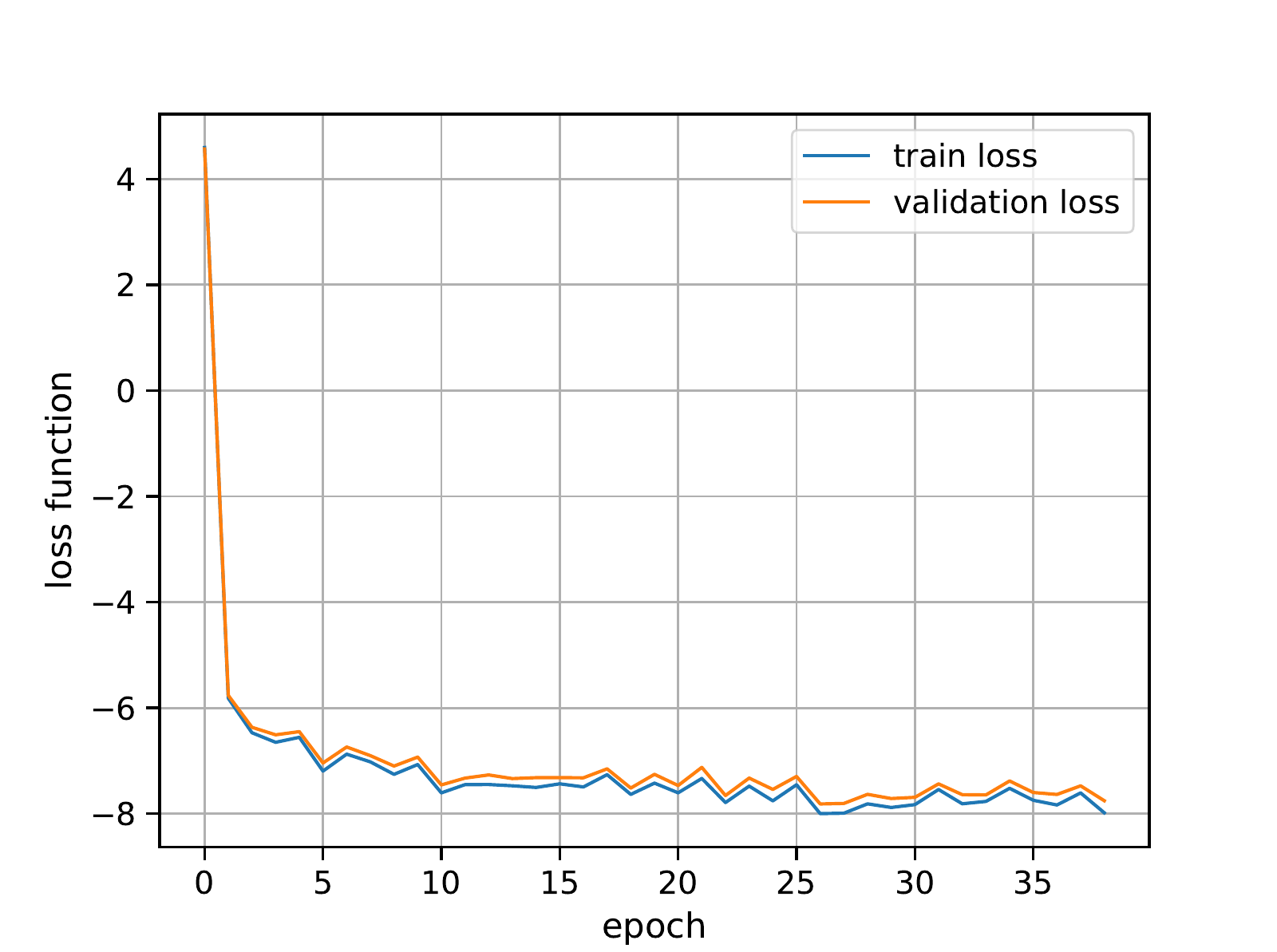}
	\caption{The value of the negative log probability of the data during the fitting of the MAF. This is an example of the learning curve of MAF, showing the change of loss function on training set and validation sets. It can be seen that the model learned quickly and converged after about 30 epochs.}
	\label{fig:lc}
\end{figure}

In all experiments, the MAF consists of 4 MADEs. We compare the results with and without DAE. For MAF without using DAE for dimensionality reduction, the hidden layers are adjusted according to $N$ (i.e. the dimensionality of $\Hb$), and are shown in Table \ref{tab:MAFconfiguration}. 
For MAF with DAE, we use DAE to reduce the dimensionality to 10 (i.e. the dimensionality of $\yb$), and the MAF has two hidden layers with 20 neurons each.  Batch normalization \citep{BatchNorm} is used between each two layers in both cases. Each time, the neural network is trained on the generated dataset with the Adam optimizer \citep{Adam} with a learning rate of $6\times 10^{-4}$ and a batch size of 30. 
During training, 80\% of the data are randomly selected to train the model, while the rest are used as validation set to evaluate the performance of the model after each epoch. To avoid overfitting, training is stopped when the performance on the validation set does not improve (i.e. the loss on the validation set does not decrease) in 30 epochs, and the epoch with the lowest loss is restored. Fig. \ref{fig:lc} shows an example of the change of the loss function during training.

When constraining the three free parameters $H_0, \Om, \Ol$ in the non-flat \La CDM model, we assume a broad uniform prior of $H_0\in [40, 100],\ \Om\in[0, 1],\ \Ol\in [0, 2]$. Following the sequential procedure proposed in Section \ref{sec:MAF_infer}, we train MAF three times in order to find the suitable range of $q(\tb)$ from which $\tb$ of training data are sampled. Each time the MAF is trained on 8000 simulated data. Because we found that the likely regions for some examples are larger than the prior range, we have to start with a $q(\tb)$ that is larger than the prior, in order to avoid fitting densities that \replaced{has}{have} hard boundaries (as discussed in Section \ref{sec:MAF_infer}). Thus, in the first time $\tb$ is sampled in a range of $H_0\in [40, 100],\ \Om\in[-1, 1],\ \Ol\in [-2.5, 2]$, which sufficiently encloses even the largest possible likely region of $P'(\tb)P(\Hb\obs|\tb)$ (the "posterior" extended to the outside of prior range, as discussed in Section \ref{sec:MAF_infer}) in our experiments. In the second and third time, the range is chosen to sufficiently enclose 10,000 samples from the first and second MAF respectively. Finally, samples are generated from the last MAF and those outside the prior is discarded. The resulting samples are the posterior estimated using MAF.

When DAE is used, after training MAF twice to find a suitable range of $q(\tb)$, a DAE is trained on the $q(\tb)$ chosen for the last MAF in the sequential procedure. It is trained using 8000 data $\{\tb, \Hb\}$ with a batch size of 32, then the encoder is used to transform the 8000 data into $\{\tb, \yb\}$,
which is used to train the last MAF. We use fully-connected networks as encoders and decoders of the autoencoders. The hidden layers of the DAEs are shown in Table \ref{tab:ae}, and the DAEs are trained using Adam optimizer \citep{Adam}  with a learning rate of $6\times 10^{-4}$. 25 \% of the data are used for validation, and training is stopped when there is no evidence of improvement on the validation set for 30 epochs.


\begin{table}
	\caption{The structure of DAE. The hidden layers of the encoder is shown in the table, and those of the decoder is simply the reverse. Data dimension $N$ refers to the number of $\{z_i, H_i, \sigma_i\}$ observed, i.e. the dimension of $\Hb$. }
	\label{tab:ae}
	\centering
	\begin{tabular}{ccc}
		\hline\hline
		Data dimension $N$ & Encode layers & 
		Encoding dimension
		\\
		\hline
		300& 200, 100 & 10 \\
		200& 200, 100 & 10 \\
		100& 100, 100 & 10 \\
		50& 40, 40 & 10 \\
		30& 25, 25 & 10 \\
		\hline
	\end{tabular}
\end{table}

For comparison, the traditional MCMC method is also used to constrain parameters from simulated OHD data. The MCMC method is carried out with Python module \lstinline|PyMC3| \citep{pymc3}, and samples are generated with No-U-Turn Sampler (NUTS) \citep{nuts}, the next-generation MCMC sampling algorithm. As pointed out in \citet{pymc3}, although there are standard sampling methods like adaptive Metropolis-Hastings and adaptive slice sampling, NUTS is the most capable algorithm provided in \lstinline|PyMC3|.
Two chains are run, both generating 50,000 samples after more than 100,000 steps of tuning (or called "burn-in"). 

\

\begin{figure*}[h]
	\centering
	\includegraphics[width=0.8\linewidth]
	{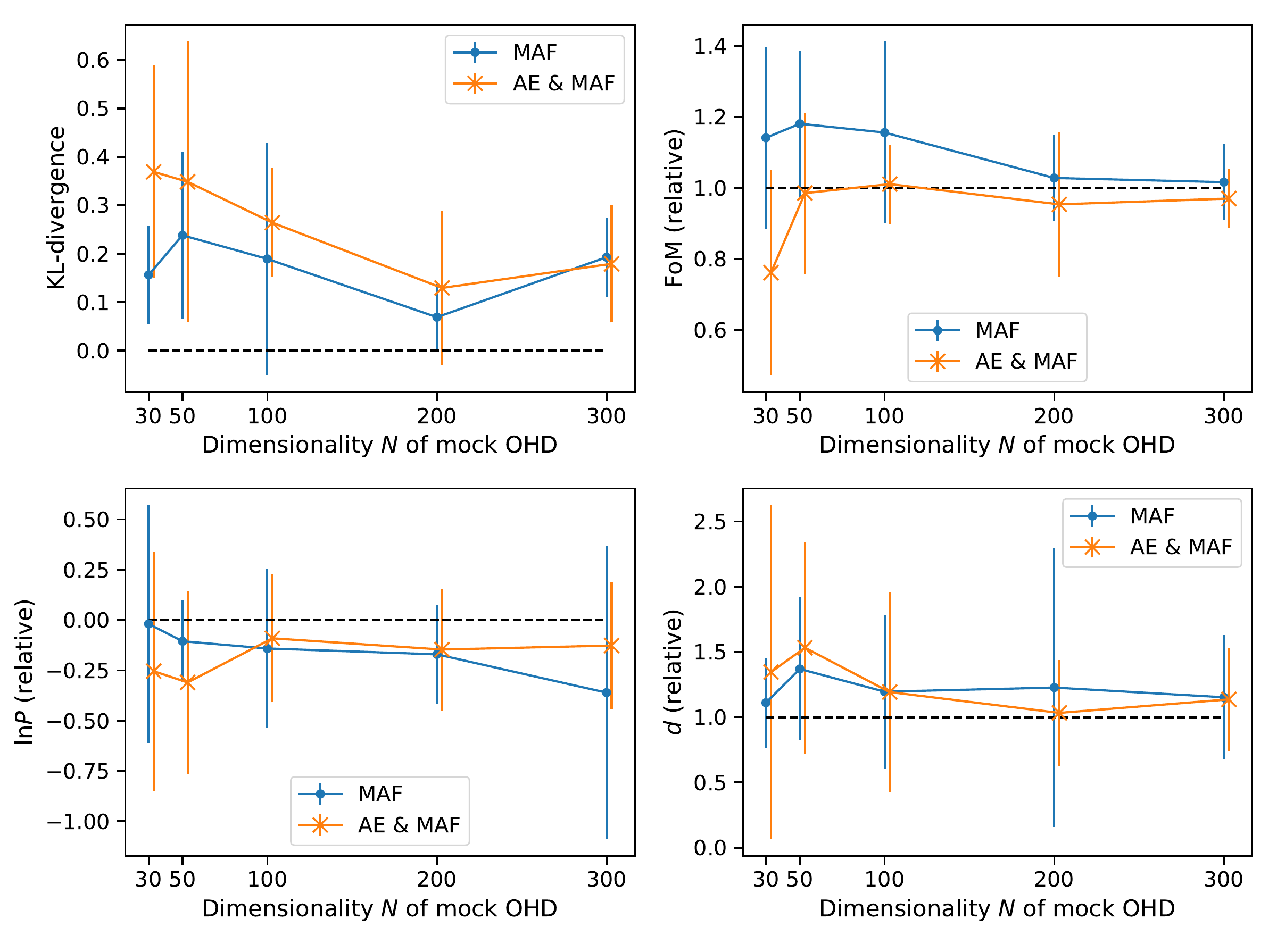}	
	\caption{The comparison of the posterior of MAF (with and without DAE) and MCMC, showing that MAF performs equally well for different dimensionality of mock data, and DAE has an insignificant effect on the posterior. The criteria used for comparison are (1) the KL divergence from the posterior of MCMC to MAF (with and without DAE), (2) figure of merit (FoM), (3) $\ln P(\tb_{\mathrm{real}}|\Hb\obs)$, and (4) distance $d$ of MAP point from real parameter. For the last three, the values of MAF relative to MCMC (as defined in Section \ref{sec:goodness}) are shown. For reference, the standard values of MCMC (relative to itself, thus they are either 0 or 1, depending on the definition) are plotted with dashed lines, so the result of MAF is the same as MCMC if a point is exactly on the dashed lines. Each data point shows the mean value and the standard deviation of the criterion, and is evaluated on the same 8 sets of mock OHD. To avoid overlap, the data with autoencoder is translated right. 
	}
	\label{fig:compare}
\end{figure*}

Both methods are used to constrain parameters from non-flat \La CDM, with free parameters $H_0, \Om, \Ol$. 
After computation, the results of MAF (with and without DAE) are compared with those of MCMC using the criteria defined in Section \ref{sec:goodness}. However,
there are two issues that worth noting. First, although $\ln P$ 
can be readily evaluated from MAF, it is not the case for MCMC method, where the probability has to be evaluated from samples with KDE. Considering the errors of the criteria for comparison that may be brought by KDE, in order to validate the comparison, a set of 100,000 samples $\tb$ is drawn from the posterior modeled by MAF  $P(\tb|\Hb\obs)$ and $\ln P$ was also estimated by KDE. 
Second, it is obvious that even if mock OHD datasets used for both models are generated from the same cosmological parameters and of same dimensionality $N$, the goodness of constraints still depends on the Gaussian noise added. 
Thus, for each $N$, all methods are evaluated on the same set of 8 mock OHD sets, which are generated with the same 8 random seeds,
then the average and standard deviation of the 8 relative performances are calculated. The results are shown in Fig. \ref{fig:compare}.


%

As can be seen in Fig. \ref{fig:compare}, MAF generally constrained cosmological parameters as well as MCMC in all of the 4 criteria for 
different $N$,
demonstrating the ability of MAF to obtain the similar posterior to that of traditional methods. For larger $N$, the likely region of posterior is smaller, but the 4 criteria of posterior are still close to that of MCMC, even though all the experiments in Fig. \ref{fig:compare} start with the same broad proposal distribution $q(\tb)$, and are trained on datasets of the same size. Thus, with the sequential method, one can get reliable estimations of the posterior with reasonable number of simulations even if the proposal $q(\tb)$ is much larger than the posterior.



The addition of the DAE has an 
insignificant
effect on the 4 criteria for mock OHD, which confirms that DAE with the loss constructed in Section \ref{sec:ae} successfully extracted most of the information contained in data about parameters. By requiring the decoder to reconstruct $\Hb$ from $\yb$ and the encoder to output $\yb$ with less variance, DAE is able to represent the information of $\Hb$ with lower-dimensional encodings $\yb$. Thus, autoencoders make it easier to train MAF and reduce computational resource occupation while obtaining a satisfactory inference of parameters.

\begin{figure*}
	\centering
	\includegraphics[width=0.8\linewidth]{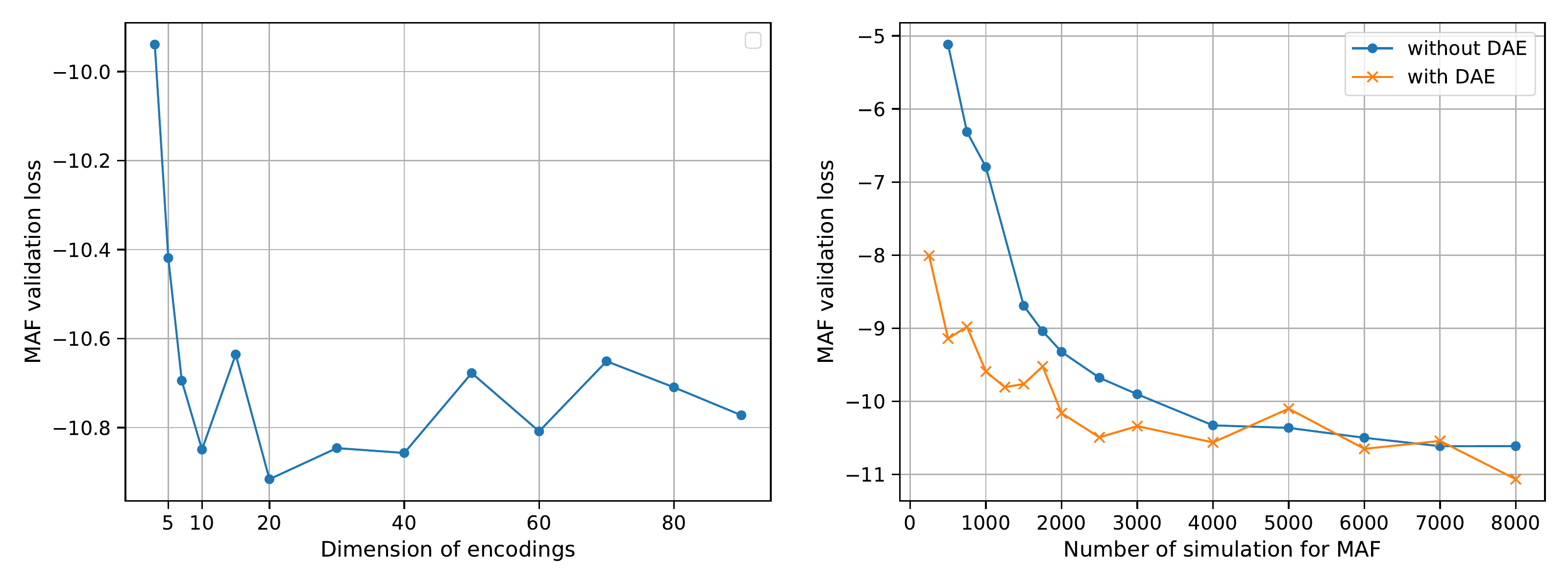}
	\caption{When the dimension of data $N=300$, the inference get stable results with at least 4000 simulation and 10-dimensional encodings, and using DAE makes MAF fit better for small number of simulations. 
	Specifically, the figure shows the change of MAF validation loss with the dimension of encodings or the number of simulation when the dimension of data $N=300$. Left panel: with 8000 simulations, the MAF validation loss decreases with dimension of encodings and get stable for dimension larger than 10. Right panel: when dimension of encodings is 10, the MAF validation loss decreases with  number of simulation for MAF with DAE (denoted with crosses) or without DAE (denoted with dots), showing that MAF fits the data better when DAE is used and tends to get stable results for larger than 4,000 simulations.}
	\label{fig:convergence}
\end{figure*}

In addition to the posterior comparison, we also explored the effect of number of simulations and the dimension of encodings on the validation loss of MAF, which represents how well the MAF fits the distribution of simulated data. We used the aforemensioned experiments with $N=300$ as an example, and repeated the last training of DAE and MAF in the sequential procedure. We changed the dimension of $\yb$ or the size of the simulated dataset, with other settings unchanged. 
As shown in Fig. \ref{fig:convergence}, 
the validation loss of the MAF decreases with the dimension of encodings $\yb$, confirming that with larger-dimensional encodings, there is less information loss in the dimensionality reduction step and MAF fits the compressed dataset better. The validation loss becomes stable for dimensions larger than 10, indicating that at least 10 summaries for 300-dimensional data are needed to make MAF learn the distribution well. Also, for MAF with or without DAE, the validation loss decreases with the number of simulations, so the more simulations, the better the MAF learns the distribution. For number of simulations larger than 4000, the validation loss tends to be stable, thus more than 4000 simulations are needed for $N=300$. When DAE is used, the MAF performs better than not using DAE, especially when the number of simulations is small. This suggests that when data is high-dimensional and expensive to simulate, using DAE makes MAF more efficient.



Finally, it is worth noting that although a DAE apparently reduces the noise level of data $\Hb$ (e.g. the square root of reconstruction MSE of DAE can be as small as $5\times 10^{-2}$ for $H$, but the typical sigma of mock $H$ data was about 20), the noise is not really removed but is transformed to some non-linear remapping of the noise, which may have a complex distribution. For some machine learning tasks, e.g. image denoising, reducing the noise level is sufficient, but in the sense of parameter constraining the data is not actually denoised. As discussed in Section \ref{sec:ae}, an encoder $f_e$ maps data $\Hb$ to the encodings $\yb$, i.e. $\yb = f_e(\Hb)$, and then MAF was trained to learn $P(\tb|\bm{y})$ rather than $P(\tb|\Hb)$. No matter how close the Hubble parameter $\Hb'$ decoded from $\bm{y}$ is with the noise-free data $\Hb\fid$, $\bm{y}$ or $\Hb'$ is just another representation of $\Hb$ with low-level complex noise, rather than an OHD with Gaussian noise that has smaller uncertainties $\sb$. The information contained in OHD cannot be increased with encoding and decoding. 




\section{Constraints with Real OHD}
\label{sec:OHD}

\begin{figure}
	\centering
	\includegraphics[width=1\linewidth]{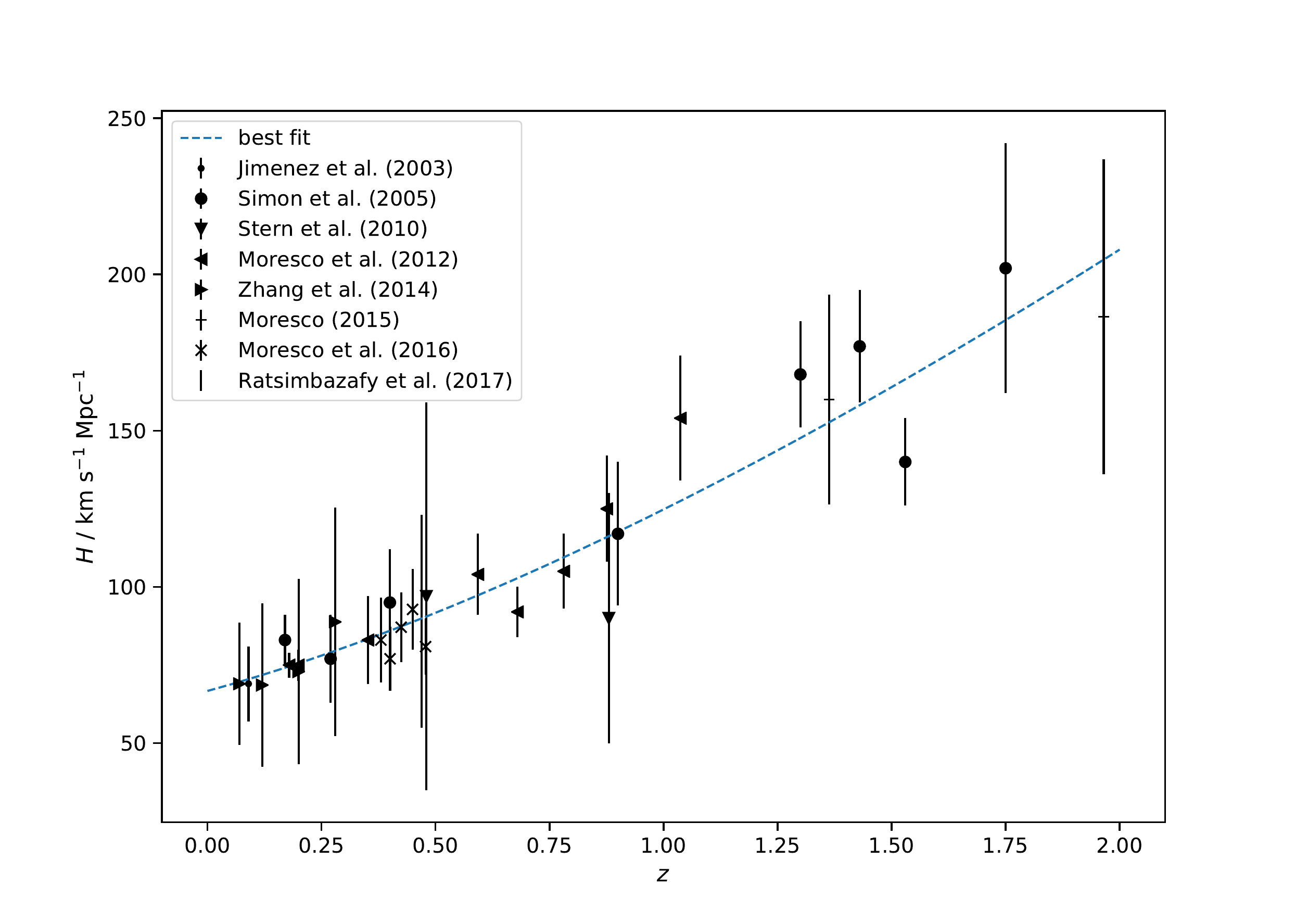}
	\caption{Current 31 Hubble parameter data points from cosmic chronometer method. The best fit found with MAF is plotted with the dashed line.}
	\label{fig:OHD}
\end{figure}

The real 
Hubble parameter data is usually measured with cosmic chronometer \citep[proposed in ][]{Jimenez2002} or from baryon acoustic oscillations (BAO) \citep{2009MNRAS.399.1663G, 2012MNRAS.425..405B, 2012MNRAS.426..226C, 2013MNRAS.431.2834X, 2013MNRAS.429.1514S, 2013AA...552A..96B, 2014MNRAS.439...83A, 2014JCAP...05..027F, 2015AA...574A..59D}. However, the determination of $H(z)$ from BAO is based on $\mathrm{\Lambda}$CDM model, making it model dependent (see  e.g. \citet{Busti2014}, \citet{Leaf2017}).
In this work, we used the 
31 OHD data evaluated with the cosmic chronometer method to constrain parameters.
The data are given in 
\citet{2003ApJ...593..622J}, \citet{2005PhRvD..71l3001S}, \citet{2010JCAP...02..008S}, \citet{2012JCAP...07..053M}, \citet{2014RAA....14.1221Z}, \citet{2015MNRAS.450L..16M}, \citet{2016JCAP...05..014M} and  \citet{2017MNRAS.467.3239R}, and are shown in Table \ref{tab:OHDCC} and Fig. \ref{fig:OHD}. 

\begin{table}
\begin{center}
\caption{31 $H(z)$ measurements from the 
cosmic chronometer
 method. $H$ and $\sigma$ are in unit of km $\rm s^{-1}$ $\rm Mpc^{-1}$.}
\label{tab:OHDCC}
\begin{tabular}{cccc}
\hline
 \emph{z} & $H$ & $\sigma$ 
  & References \\
\hline
0.09	&	$	69$&$12	$	
 &	\citet{2003ApJ...593..622J} \\
\hline
0.17	&	$	83$&$8	$	
 &	\\
0.27	&	$	77$&$14	$	
 &	\\
0.4	&	$	95$&$17	$	
 &	 \\
0.9	&	$	117$&$23	$	
 &	\citet{2005PhRvD..71l3001S} \\
1.3	&	$	168$&$17	$	
 &	\\
1.43	&	$	177$&$18	$	
 &	\\
1.53	&	$	140$&$14	$	
 &	\\
1.75	&	$	202$&$40	$	
 &	\\
\hline
0.48	&	$	97$&$62	$	
 &	\citet{2010JCAP...02..008S} \\
0.88	&	$	90$&$40	$	
 &	\\
\hline
0.1791	&	$	75$&$4	$	
 &	\\
0.1993	&	$	75$&$5	$	
 &	\\
0.3519	&	$	83$&$14	$	
 &	\\
0.5929	&	$	104$&$13	$	
 &	\citet{2012JCAP...07..053M} \\
0.6797	&	$	92$&$8	$	
 &	\\
0.7812	&	$	105$&$12	$	
 &	\\
0.8754	&	$	125$&$17	$	
 &	\\
1.037	&	$	154$&$20	$	
 &	\\
\hline
0.07	&	$	69$&$19.6	$	
 &	\\
0.12	&	$	68.6$&$26.2	$	
 &	\citet{2014RAA....14.1221Z} \\
0.2	&	$	72.9$&$29.6	$	
 &	\\
0.28	&	$	88.8$&$36.6	$	
 &	\\
\hline
1.363	&	$	160$&$33.6	$	
 &	\citet{2015MNRAS.450L..16M} \\
1.965	&	$	186.5$&$50.4	$	
 &	\\
\hline
0.3802	&	$	83$&$13.5	$	
 &	\\
0.4004	&	$	77$&$10.2	$	
 &	\\
0.4247	&	$	87.1$&$11.2	$	
 &	\citet{2016JCAP...05..014M} \\
0.4497	&	$	92.8$&$12.9	$	
 &	\\
0.4783	&	$	80.9$&$9	$	
 &	\\
\hline
0.47   &      $      89$&$34        $      
 &   \citet{2017MNRAS.467.3239R} \\
\hline
%
\end{tabular}\label{table:1}
\end{center}
\end{table}

\begin{figure*}[htbp]
	\centering
	\includegraphics[width=0.7\linewidth]
{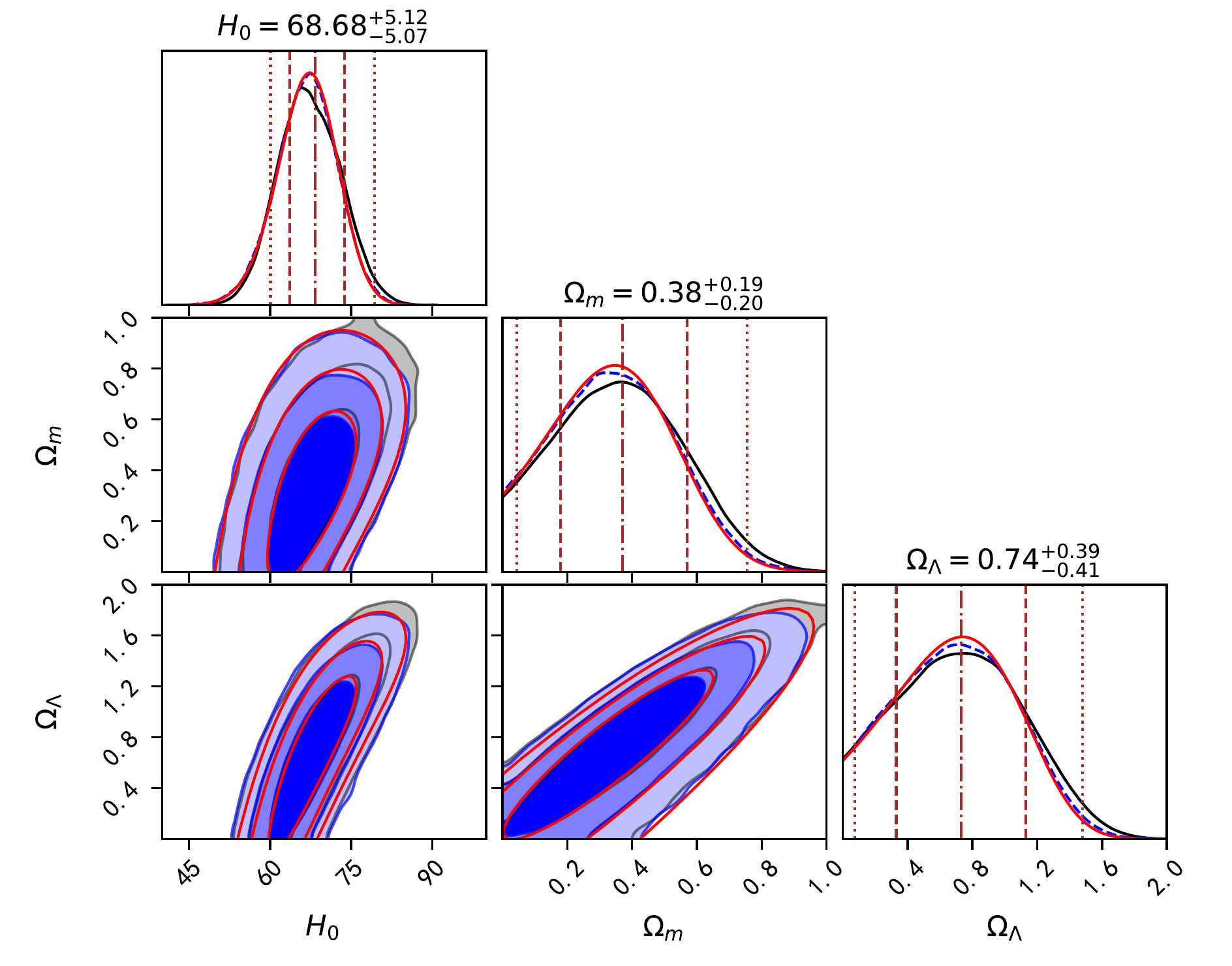}
	\caption{The posterior estimated with MAF is consistent with the MCMC and the directly calculated posterior.
$68.26\%, 95.44\%, 99.74\%$ confidence regions of posterior estimated from real OHD using MCMC (in blue) and MAF (in black) are shown. The real posterior is also calculated analytically using Eq. \ref{chi2} (in red). Both MAF and MCMC gave estimations very close to the real posterior. The $68.26\%$ confidence intervals estimated with MAF are also shown in the titles, and the $68.26\%, 95.44\%$ confidence intervals and medians of each parameter estimated with MAF are also marked on the diagonal.
	}
	\label{fig:postcomp}
\end{figure*}

Given the experiments and results in Section \ref{sec:SimRes}, the aforementioned methods, i.e. MCMC and MAF, are used to constrain cosmological parameters from real OHD data. We assume the same prior specified in Section \ref{sec:SimRes}: $H_0\in [40, 100],\ \Om\in[0, 1],\ \Ol\in [0, 2]$. The 
MAF is the same as the one in Section \ref{sec:SimRes} that was used to process 30 OHD data and is trained on 25,000 simulations. The MAF is trained once (while in Section \ref{sec:SimRes} it is trained three times using the proposed sequential procedure) with $\tb$ sampled in a range of $H_0\in [40, 100],\ \Om\in[-1, 1],\ \Ol\in [-2.5, 2]$. The posteriors of both methods, together with the posterior directly calculated from Equation (\ref{like}), are shown in Fig. \ref{fig:postcomp}. 
As can be seen, the results of MCMC are almost the same as the directly calculated one, which demonstrates the ability of MCMC to represent the real posterior, and justifies the evaluation of MAF and DAE with MCMC as a reference. The confidence regions estimated with MAF are very close to those of MCMC and the result directly calculated using the analytic expression of likelihood, again showing MAF's capability of giving satisfiable constraints without knowing the analytical form of likelihood. With MAF, we find $H_0 = 68.68 ^{+5.12}_{-5.07}\ \mathrm{km\ s^{-1}\ Mpc^{-1}},\ \Om = 0.38 ^{+0.19}_{-0.20},\ \Ol = 0.74 ^{+0.39}_{-0.41}$. This is consistent with the result using MCMC method, which gives $H_0 = 68.22 ^{+4.66}_{-4.67}\ \mathrm{km\ s^{-1}\ Mpc^{-1}},\ \Om = 0.36 ^{+0.18}_{-0.19},\ \Ol = 0.71 ^{+0.37}_{-0.39}$. 



\section{Constraints from SNe Ia}
\label{sec:SN}

Type Ia supernova (SN Ia) dataset is another widely-used dataset for cosmology research. When used to constrain cosmological parameters, observed SN Ia light curves are fitted with common light curve fitters, e.g. SALT2 \citep{Guy2007}, MLCS2K2 \citep{Jha2007}, and cosmological parameters are traditionally constrained with likelihood function and $\chi^2$ statistic of distance modulus $\mu(z)$. 
In this paper, we used SALT2, as used in \citet{Scolnic2018}. A SALT2 model fits light curves with 3 free parameters, $x_0$, the overall normalization, $x_1$, the deviation from the average shape of light curves, and $c$, the deviation from the average color (see reference such as \citet{Amanullah2010, Suzuki2012, Yang2013} for more on SALT2). Other parameters including redshift $z$ are also fitted from light curves. Then, the observational distance modulus $\mu\obs$ can be determined by 
\begin{equation}
\mu\obs = m_B - M_B +\alpha x_1 - \beta c,
\end{equation}
where $m_B$ is integrated B-band flux that can be calculated given $x_0$, $x_1$ and $c$; $M_B$ is the absolute B-band magnitude, and $\alpha,\beta$ are the coefficients of the relation between luminosity and $x_1, c$. Unfortunately, $\alpha$ and $\beta$  are usually "nuisance parameters" that have to be decided before $\mu\obs$ can be calculated. 
 Thus, they are fitted simultaneously with cosmological parameters $\tb$, 
traditionally
by calculating the $\chi^2$ statistic. One simple definition of the $\chi^2$ statistic is 
\begin{equation}
\chi^2 = \sum_i \frac{[\mu_{\mathrm{obs}, i}(\alpha, \beta) - \mu(z_i; \tb)]^2}{\sigma_i^2}.
\end{equation}

However, as pointed out by \citet{Weyant2013}, the Gaussian distribution is too simple to model the uncertainties in SN Ia datasets, but it is difficult to perform traditional methods without an analytical form of likelihood. Thus, we explore the method discussed in this paper, a combination of DAE and MAF, to constrain free parameters (cosmological parameters $\tb$ and $\alpha, \beta$) from SN data\footnote{Here data $\Xb$ corresponds to the vector $\Hb$ in the OHD problem. } $\Xb\obs$. We propose performing inference directly from data $z, x_0, x_1, c$, rather than $\mu(z)$. Therefore, the real observed data $\Xb\obs$ and the each element $\Xb_i$ in the simulated training dataset $\{\Xb_i, \tb_i\}$ is an $N \times 4$ matrix, where $N$ is the total number of real observed SN Ia, and the column of the matrix represents $z, x_0, x_1, c$ respectively.\footnote{
The $z, x_0$ data used to train ANNs was actually $\ln z$ and $\ln x_0$ in this work, but we still denote them as $(z, x_0, x_1, c)$ to highlight their physical meanings.} Since each $\Xb_i$ is regarded as one instance of data, the dimensionality of data is as large as $4N$, which necessitates demensionality reduction with DAE. 

The simulation model simulates the process of observing light curves and performing light curve fitting with SALT 2, which is implemented with the supernova analyzing package SNANA\footnote{SNANA v10\_76b.} \citep{Kessler2009a}. To generate simulated training data for neural networks, free parameters were first sampled from a uniform prior. Given each group of free parameters, $N$ light curves are first simulated and then fitted 
to get data $\Xb_i$. When simulating SN light curves, rest-frame light curves are generated according to the model (SALT2 in this paper), then host-galaxy extinction, K-correction (transforming UBVRI to observer-frame filters), Galactic extinction are applied. Finally, the magnitudes of light curves are translated into CCD counts accounting for atmospheric transmission and telescope efficiency. In the fitting step, the light curves are fitted with the chosen SN model to get parameters such as $z, x_0, x_1, c$.\footnote{The simulation and fitting of light curves are made using the programs \lstinline|snlc_sim.exe| and \lstinline|snlc_fit.exe| of SNANA respectively.} (See \citet{Kessler2009a} and the manual of SNANA\footnote{Available at \url{https://snana.uchicago.edu/}.} for detailed information about the algorithms.) 

Note that the simulation model described above 
includes both simulation and fitting of light curves, and
is much more complicated than that of OHD in this paper, thus the traditional $\chi^2$ method cannot be applied due to the absence of a tractable likelihood.
After the simulation, DAE and MAF can be trained to estimate free parameters given observational SN Ia data. Since the real noise-free data of $z, x_0, x_1, c$ are intractable, the DAE was trained to reconstruct data simulated from SNANA simulating program, given those fitted from simulated light curves as inputs. This reduces the noise added in the process of light curve fitting.

In this paper, we carried out a preliminary constraint on cosmological parameters in non-flat \La CDM model to demonstrate the application of our procedure to SNe Ia. We used 
the latest compilation of SN Ia data called Pantheon \citep{Scolnic2018},
which consists of 1048 SNe from several different surveys, namely SDSS, SNLS, HST, Low-$z$ and PS1. Thus, when simulating $N=1048$ data in each iteration, we used different simulation configuration files corresponding to these surveys to generate mock Pantheon data that look like real Pantheon. In addition, the same cut as in Table 2 of \citet{Scolnic2018} was used so that the mock data all passed these cut criteria.

Note that less prior knowledge and assumptions are made in this paper than some similar works. In \citet{Scolnic2018}, $\alpha$ and $\beta$ were determined with BEAMS with Bias Corrections (BBC) method proposed by \citet{Kessler2017}. However, the BBC method requires an assumption of cosmological model and reference values (or prior) of parameters, making nuisance parameters model dependent. Thus, BBC was not used in this paper. In addition, in \citet{Weyant2013}, ABC was used to perform likelihood-free inferences, but inference was made using $\mu(z)$ data rather than $(z, x_0, x_1, c)$. As mentioned above, distributions of nuisance parameters are needed to get $\mu(z)$, and in \citet{Weyant2013} the nuisance parameters were drawn from empirical distributions. However, the data directly fitted from light curves used in this paper are more model independent, which do not necessarily need assumptions of cosmological models or priors of parameters.

According to the posteriors of parameters given in \citet{Scolnic2018}, we set the uniform prior $H_0 \in [50, 90],\ \Om \in [0, 0.7],\ \Ol \in [0.2, 1.3],\ \alpha \in [0.12, 0.21],\ \beta\in [2.54, 4.15]$ so that they cover the whole possible region of the posterior. Then about 18,000 samples of parameters were drawn, and corresponding training data $\Xb$ was simulated with SNANA. In our experiments, 
each $\Xb_i$ was standardized into $[0,1]$ along the columns before training. Following the hyperparameters in Section \ref{sec:simul}, we constructed a DAE model with an encoder that has three hidden layers with 1000, 500 and 250 neurons in each respectively, and trained it to extract 70-dimensional encodings from the data with a learning rate of $8\times 10^{-5}$. Then, a MAF with 3 MADEs, each of which has 2 hidden layers with 75, 50 neurons respectively, was trained to estimate the 5-dimensional free parameter $(H_0, \Om, \Ol, \alpha, \beta)$, with a learning rate of $7\times 10^{-4}$. After the training, the posterior was estimated by feeding the real Pantheon data (standardized the same way as training data) to the ANNs. 

\begin{figure*}[ht]
	\centering
	\includegraphics[width=0.7\linewidth]
{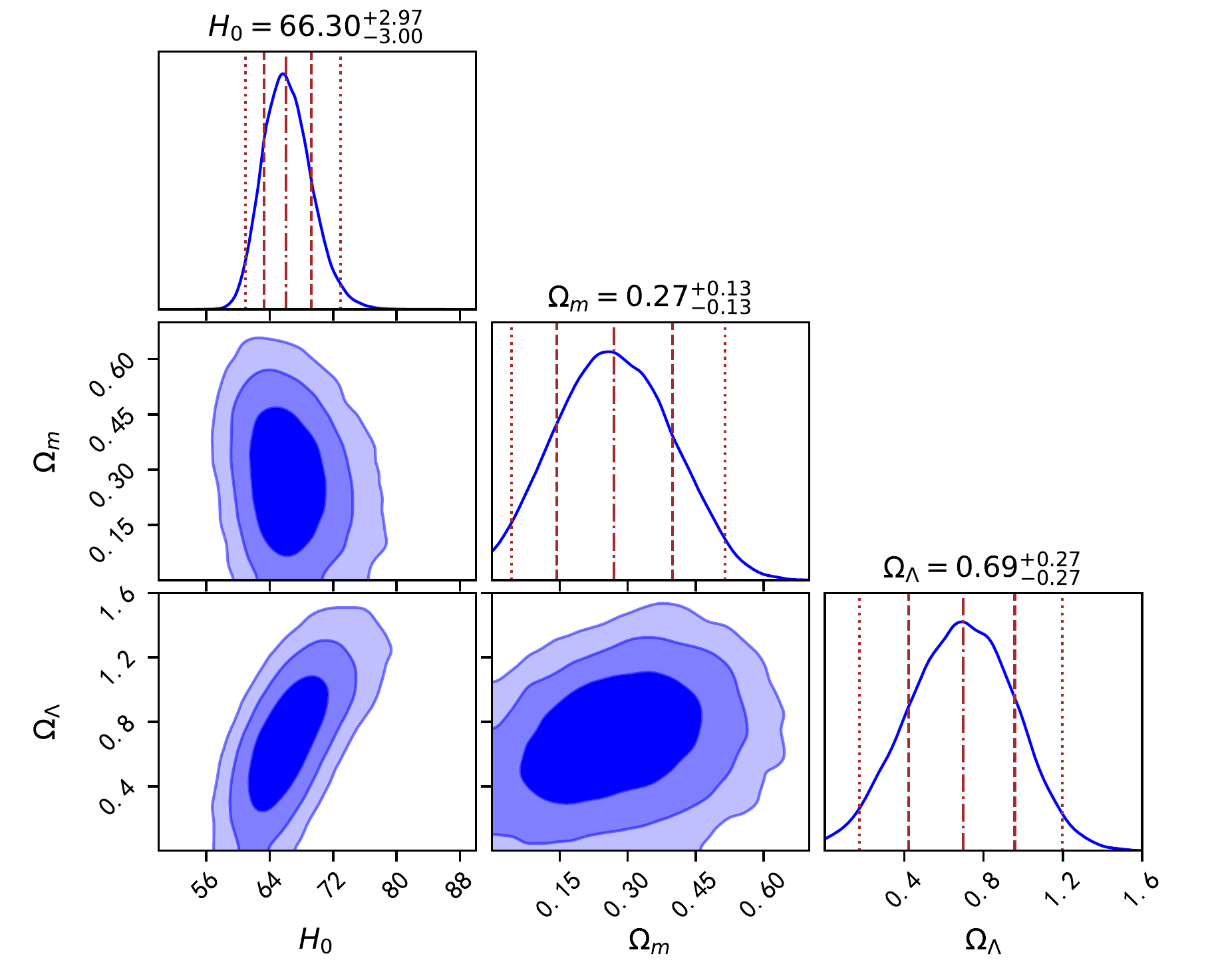}
	\caption{The posterior estimated with DAE and MAF using Pantheon data ($\alpha, \beta$ are marginalized) and the methods in this paper. The $68.26\%, 95.44\%, 99.74\%$ confidence regions are plotted for 2D marginal distributions; The median and $68.26\%, 95.44\%$ confidence interval are marked for 1D marginal distributions. }
	\label{fig:SN}
\end{figure*}

The constraint of cosmological parameters is shown in Fig. \ref{fig:SN}, where nuisance parameters $\alpha$, $\beta$ are marginalized out. We find $H_0 = 66.3 \pm 3.0\  \mathrm{km\ s^{-1}\ Mpc^{-1}},\ \Om = 0.27 \pm 0.13,\ \Ol = 0.69 \pm 0.27$, and nuisance parameters $\alpha = 0.149\pm 0.024,\ \beta = 3.27 \pm 0.38$. The constraint is similar to the value of non-flat \La CDM constraint in \citet{Scolnic2018} of $\Om = 0.319 \pm 0.070, \Ol = 0.733 \pm 0.133$, and the BBC result $\alpha = 0.167 \pm 0.012$, $\beta = 3.51 \pm 0.16$ (C11 scatter model), $\beta = 3.02\pm 0.12$ (G10 scatter model), but the confidence region is larger. This may be the result of uniform priors of cosmological parameters and nuisance parameters; with more stringent prior knowledge, the confidence region will be smaller. Another reason is the more degree of freedom in the model considered in this paper, for there are 3 more free parameters than the non-flat \La CDM in \citet{Scolnic2018}, i.e. $H_0$, $\alpha$ and $\beta$. In addition, the constraint is expected to be more precise than the result in Fig. \ref{fig:SN} with more accurate simulation model and larger training set. 
Thus, one has to choose hyperparameters to balance accuracy and efficiency in application.








\section{Conclusions and Discussion}
\label{sec:concl}

In this paper, it is proposed that two kinds of ANNs, DAE and MAF, can be used together to perform likelihood-free cosmological constraints, 
as long as there is a model to simulate observational data. Our procedure used DAE for the first time to extract features from data before estimating the posterior of parameters with MAF, so that each MADE can be less complex, and more MADEs is possible to be stacked in MAF. An additional term of loss function besides from the traditional MSE was proposed to train the DAE for likelihood-free inference tasks.
A sequential training procedure, modified from \citet{Papamakarios2019}, was also adopted for problems where the likely region of the posterior is unknown.
The 
proposed procedures
were tested on Hubble parameter $H(z)$ and Pantheon SN Ia datasets. For $H(z)$ data, simple Gaussian uncertainties were assumed due to the absence of a sophisticated OHD simulation model, and Gaussian process was used for the first time to estimate the standard deviation $\sigma(z)$ of data. The ANN and MCMC method were used to contrain parameters from mock $H(z)$, and were evaluated on several performance criteria, including KL divergence and the commonly used FoM and negative log probability of true parameters. When constraining cosmological parameters from a non-flat \La CDM model,
the MAF with and without DAE appeared to achieve performance comparable to MCMC. The MAF gets goodness of performance that has negligible difference from the traditional MCMC method; adding DAE does not significantly reduce the performance, but reduces the dimensionality of data, making the MAF smaller and learn the distribution better with limited number of simulations. 

The procedure proposed in this paper was applied to real OHD and SN Ia data. MAF was used to give constraints
from real observational $H(z)$ data (OHD), and we got $H_0 = 68.68 ^{+5.12}_{-5.07}\ \mathrm{km\ s^{-1}\ Mpc^{-1}},\ \Om = 0.38 ^{+0.19}_{-0.20},\ \Ol = 0.74 ^{+0.39}_{-0.41}$, which is consistent 
to the result of MCMC.
For SN Ia data, we proposed constraining cosmological parameters together with nuisance parameters $\alpha, \beta$ directly using data $z, x_0, x_1, c$ fitted from light curves, with less necessary prior knowledge required. This is a combination and extension of the treatments in \citet{Weyant2013} and \citet{Scolnic2018}. Using MAF and DAE in this paper and a simulation model implemented with the SNANA package, we got a preliminary constraint on the non-flat \La CDM model with Pantheon dataset: $H_0 = 66.3 \pm 3.0 \mathrm{\ km\ s^{-1}\ Mpc^{-1}},\ \Om = 0.27 \pm 0.13,\ \Ol = 0.69 \pm 0.27$, which is similar to the results in \citet{Scolnic2018}.

Likelihood-free inference is an important method when estimating cosmological parameters from simulation-based models, where likelihood is not as simple as the Gaussian distribution or even intractable. Although the uncertainty model for $H(z)$ in this paper is Gaussian, the DAE and MAF do not use the analytical expression of the Gaussian likelihood. In future, it is advised that more sophisticated simulation models for OHD be established so that more realistic simulated data can be generated, which enables better treatment of the uncertainties of OHD. Such a simulation model is expected to simulate the spectrum of luminous red galaxies as well as the process of spectral fitting and the cosmic chronometer method. The model is supposed to be
 even 
better if a more accurate galaxy spectrum simulation and fitting model is 
adopted, 
e.g. the Yunnan evolutionary population synthesis models \citep{Li2008,Zhang2012,Li2013,Moresco2018}. Apart from cosmological parameters, MAF and DAE can also be used to estimate other parameters from galaxy spectrums, such as the age and metallicity. In addition, since the MAF can give estimations as long as an accurate simulation model is established, it has the potential to be applied to other datasets. For example, cosmological parameters can be constrained from the large-scale structure by adding convolutional properties to the ANNs
used in this paper, following \citet{Pan2019a}.


Admittedly, the combination of DAE and MAF in this paper is not perfect in some aspects. Compared to the simulation-based MCMC method, an accurate estimation of the posterior might require more MADE layers,
which means the need for a larger training dataset. However, it may be time consuming and computationally expensive for very slow or complex simulation models to generate enough training data. Also, ANN models usually need careful hyperparameter fine-tuning before the best performance can be achieved. Therefore, future work should be focused on better models or strategies to perform faster and more accurate constraints.







\section*{Acknowledgements}
We thank the anonymous referee for the comments that helped us greatly improve this paper. We thank Shu-Lei Cao, Wei Liu, Jin Qin, Jing Niu and
 Yun-Long Li 
for useful discussions and thank Kang Jiao, Che-Qiu Lyu and Cheng-Zong Ruan for their kind help. This work was supported by the National Science Foundation of China (Grants No.11929301), and National Key R\&D Program of China (2017YFA0402600). 




\bibliography{library,OHD_ref,myref1}
\bibliographystyle{aasjournal}

\end{document}